\documentclass[aps,prd,twocolumn,superscriptaddress,amsmath,preprintnumbers,amssymb,nofootinbib,longbibliography]{revtex4-2}

\usepackage{graphicx}
\usepackage[dvipsnames]{xcolor}
\usepackage{bbm}
\usepackage[colorlinks]{hyperref}
\usepackage[normalem]{ulem}

\usepackage{tikz}
\usetikzlibrary{tikzmark}
\usetikzlibrary{positioning}

\hypersetup{linkcolor=BrickRed,citecolor=Green,
filecolor=Mulberry,
urlcolor=NavyBlue,
menucolor=BrickRed,
runcolor=Mulberry
}

\newcommand{\newtext}[1]{{\color{black}{#1}}}

\definecolor{lime}{HTML}{A6CE39}
\DeclareRobustCommand{\orcidicon}{\hspace{-1mm}
	\begin{tikzpicture}
	\draw[lime, fill=lime] (0,0) 
	circle [radius=0.16] 
	node[white] {{\fontfamily{qag}\selectfont \tiny \,ID}};
	\draw[white, fill=white] (-0.0525,0.095) 
	circle [radius=0.007];
	\end{tikzpicture}
	\hspace{-3mm}
}

\foreach \x in {A, ..., Z}{\expandafter\xdef\csname orcid\x\endcsname{\noexpand\href{https://orcid.org/\csname orcidauthor\x\endcsname}
			{\noexpand\orcidicon}}
}

\allowdisplaybreaks

\begin{document}
\preprint{N3AS-24-035}

\title{Quantum Closures for Neutrino Moment Transport}

\author{James P. Kneller\orcidA{}}
\email{jim\_kneller@ncsu.edu}
\affiliation{Department of Physics, North Carolina State University, Raleigh, NC 27695, USA}

\author{Julien Froustey\orcidB{}}
\email{jfroustey@berkeley.edu}
\affiliation{Department of Physics, North Carolina State University, Raleigh, NC 27695, USA}
\affiliation{Department of Physics, University of California Berkeley, Berkeley, CA 94720, USA}

\author{Evan B. Grohs\orcidE{}}
\affiliation{Department of Physics, North Carolina State University, Raleigh, NC 27695, USA}

\author{Francois Foucart\orcidF{}}
\affiliation{Department of Physics \& Astronomy, University of New Hampshire, Durham, NH 03824, USA}

\author{Gail C. McLaughlin\orcidC{}}
\affiliation{Department of Physics, North Carolina State University, Raleigh, NC 27695, USA}

\author{Sherwood Richers\orcidD{}}
\affiliation{Department of Physics, University of Tennessee Knoxville, Knoxville, TN 37996, USA}

\begin{abstract}

A computationally efficient method for calculating the transport of neutrino flavor in simulations is to use angular moments of the neutrino one-body reduced density matrix, i.e., `quantum moments'. As with any moment-based radiation transport method, a closure is needed if the infinite tower of moment evolution equations is truncated. We derive a general parameterization of a quantum closure and the limits the parameters must satisfy in order for the closure to be physical. We then derive from multi-angle calculations the evolution of the closure parameters in two test cases which we then progressively insert into a moment evolution code and show how the parameters affect the moment results until the full multi-angle results are reproduced. This parameterization paves the way to setting prescriptions for genuine quantum closures adapted to neutrino transport in a range of situations.
\end{abstract}
\maketitle

\section{Introduction}

Over the past decade, a large body of evidence has accumulated that indicates neutrino flavor evolution occurs interior to the shock in a core-collapse supernova, and in the vicinity of the accretion disk in a compact object merger — for reviews of this evidence and the multiple aspects of flavor evolution involved we refer the reader to \cite{Duan:2010bg,Mirizzi:2015eza,Tamborra:2020cul,Capozzi:2022slf,Richers_review,Volpe:2023met,Sen:2024fxa}. Flavor transformation in these locations will alter the dynamics of the system~\cite{Pejcha:2011en,Stapleford:2019yqg,2023PhRvD.108d3006S,Nagakura:2023mhr,Ehring:2023abs,Nagakura:2024trv,Xiong:2024tac}, the nucleosynthesis that occurs~\cite{McLaughlin:1999pd,Duan:2010af,Malkus:2012ts,Wu:2014kaa,Zhu:2016mwa,Sasaki:2017jry,Wu:2017drk,Xiong:2020ntn,George:2020veu,Li:2021vqj,Fujimoto:2022njj,Fischer:2023ebq,Wang:2023tso,Balantekin:2023ayx} and the expected neutrino signal, e.g.~\cite{Fuller:1998kb,2002astro.ph..5390S,2008PhRvD..77d5023K,Gava:2009pj,Caballero:2009ww,2011JCAP...10..019V,Patton:2014lza,2015PhRvD..91f5016W,2017PhRvD..96j3008W,2018MNRAS.480.4710S,2024JCAP...03..040M}. However flavor transformation is typically not accounted for in the simulations of core-collapse supernovae and/or compact object mergers because the computational expense is expected to increase by several orders of magnitude due to the much finer spatial fidelity and the concomitant decrease in time step sizes. In addition, if the transport is solved using discrete ordinates, the angular resolution of the neutrino radiation field will likely need to increase by a factor of several hundreds or thousands \cite{2012PhRvD..86l5020S} and the often-assumed axial symmetry will need to be dropped \cite{2013PhRvL.111i1101R,2013PhRvD..88g3004M,2015PhRvD..92j5020M,Shalgar:2021oko}, thus necessitating inclusion of additional spatial coordinates and propagation directions. At the present time such calculations are not feasible. 

It has been suggested that the huge increase in the computational expense of discrete ordinate methods due to the necessary angular resolution for convergence of results may be mitigated by using angular moments~\cite{Thorne:1981nvt,Shibata:2011kx,Cardall:2012at} of the radiation field. In order to include flavor transformation effects, the moments need to be generalized by using a density matrix to describe the distribution of the system in the Hilbert space rather than a scalar distribution function. We call such generalizations `quantum moments'. 

\newtext{The generalization to quantum moments does not do away with the two fundamental issues of all moment approaches to transport: at which order to truncate the tower of moment evolution equations, and then what to do with the truncated moments that appear in the remaining evolution equations. Two-moment schemes are a popular choice as they seem to occupy the sweet spot between accuracy and complexity in many applications. Determining how to truncate the hierarchy is a subtle task, however.
When truncating the hierarchy, one might try to simply assume that all moments beyond a particular rank are zero, though this can lead to pathologies. For example, 
it has notably been shown in the context of the fast flavor instability (FFI) that there is a cascade of power from large to small angular scales~\cite{Johns:2019izj,Johns:2020qsk} as the system evolves, so that when the power in the higher moments becomes significant, the results from a scheme where these moments are set to zero will begin to diverge from the true evolution. In particular, such a choice would be pathological since the trace of any moment of rank $n$ must be equal to the moment of rank $n-2$. One could create a two-moment scheme by setting the pressure tensor (rank 2) to be zero, for example, but this would be mathematically wrong when the energy density moment (rank 0) is nonzero.
Thus it is much better to use a \emph{closure}: a prescription for constructing the higher-order moments as a function of the lower-order ones. If the closure is chosen well — or we are lucky — the results from the moment calculation will match those from a multi-angle calculation.}
The closure problem has been long studied in classical neutrino transport, with a number of prescriptions, usually analytic or based on multi-angle calculations~\cite{Minerbo_1978,Levermore_1984,Cernohorsky_closure_1994,Smit_closure_2000,Rampp:2002bq,Muller:2010ymw,Murchikova:2017zsy,Foucart:2017mbt,Richers:2020ntq}. Up to now, dynamical quantum calculations with moments have been performed using ad hoc, semi-classical closures: scalar prescriptions in~\cite{Myers:2021hnp}, or semi-classical extensions of the Minerbo closure based on the flavor trace of the different moments in~\cite{Grohs:2022fyq,Grohs:2023pgq,Froustey:2023skf,Froustey:2024ln}. Recently, a fully quantum generalization of the Minerbo closure was derived in~\cite{Froustey_QMinerbo}, albeit limited to the case of small flavor coherence.

We highlight that the angular moments obtained in classical neutrino transport are commonly used to study quantum flavor evolution. For instance, when trying to determine whether a system is prone to a FFI, one must generally look for a crossing between the angular distributions of neutrinos and antineutrinos — a feature that is, by essence, hidden in the averaged information available in a two-moment method. There have nevertheless been a range of methods developed to extract the instability criterion from only the first angular moments, see~\cite{Richers_evaluating_2022} and references therein. However, such studies only seek to determine whether crossings are present between the (underlying) flavor off-diagonal components of the (anti)neutrino distributions, regardless of the angular properties of the flavor off-diagonal components. This is a suitable approach to determine if an instability occurs, but it leaves wide open the question of an adequate, fully quantum closure, that is required if one wants to actually \emph{evolve} the quantum angular moments.

The purpose of our study is to investigate quantum closures and to develop a framework that allows us to formulate them in a general way. As a foundation for this study we introduce the quantum angular moments and their evolution equations in Sec.~\ref{sec:EvolutionEquations}. In Sec.~\ref{subsec:parameterization_closure}, we start from a general ansatz and show how a quantum closure can be understood in terms of the eigenvalues of the moments and the unitary matrices which diagonalize the moments. For two neutrino flavors we find the closure is described by four parameters and provide two alternative parameter sets for it. We then show in Sec.~\ref{subsec:limits_params} that the parameters relating the energy density and pressure tensor moments are not arbitrary and need to satisfy several constraints in order that the closure be physical. In Sec.~\ref{sec:numerical_calculations}, we test our general framework by applying it to two test cases where we are able to compute the exact closure from multi-angle calculations. We then progressively insert the parameterization into a moment code and show how increasingly accurately closures eventually reproduce the results from the multi-angle calculations. \newtext{For the second test case, we also design a simple a priori closure prescription, whose encouraging results show great promise for future studies of more complex closures.} We summarize and conclude in Sec.~\ref{sec:conclusions}.

Throughout this paper, we use natural units in which $\hbar = c = k_\mathrm{B} = 1$.


\section{Expressing Neutrino Flavor Evolution in Moments}
\label{sec:EvolutionEquations}

The neutrinos and antineutrinos are described by one-body reduced density matrices $\varrho(t,\vec{r},\vec{q})$ and ${\bar{\varrho}}(t,\vec{r},\vec{q})$ respectively, where $t$ is time, $\vec{r}$ is the space coordinate, and $\vec{q}$ is the neutrino momentum. We shall take the number dimension of the neutrino flavor space to be $N_f=2$ when we need to be specific. 
The angular moments of the density matrix are defined following \cite{2013PhRvD..88j5009Z} to be
\begin{align}
E(t,\vec{r},q) &= \frac{1}{4\pi}\,\left(\frac{q}{2 \pi}\right)^{3} \int d\Omega_{q}\,\varrho(t,\vec{r},\vec{q})  \, , \label{eq:defE}\\
\vec{F}(t,\vec{r},q) & = \frac{1}{4\pi}\,\left(\frac{q}{2 \pi}\right)^{3}  \int d\Omega_{q}\,\widehat{q}\, \varrho(t,\vec{r},\vec{q}) \, , \label{eq:defF} \\
P(t,\vec{r},q) & = \frac{1}{4\pi}\,\left(\frac{q}{2 \pi}\right)^{3}  \int d\Omega_{q}\, \widehat{q}\otimes \widehat{q}\, \varrho(t,\vec{r},\vec{q}) \, , \label{eq:defP}
\end{align}
where $q = |\vec{q}|$. Note that the prefactor of $1/4 \pi$, included here, depends on the convention chosen.

In spherical symmetry, the evolution equations for the density matrices read, in the mean-field approximation (see, e.g,~\cite{OConnor:2012lcd}):
\begin{equation}
\label{eq:QKE_multi}
    \frac{\partial \varrho}{\partial t} + \mu \frac{\partial \varrho}{\partial r} + \frac{1 - \mu^2}{r} \frac{\partial \varrho}{\partial \mu} = - \imath \left[H, \varrho \right] \, , 
\end{equation}
with $\mu = \hat{q} \cdot \hat{r}$ the neutrino propagation angle measured from the radial direction and $H$ the Hamiltonian. We have omitted the terms associated with emission, absorption and scattering. 
The mean-field Hamiltonian is written $H = H_V + H_M + H_{SI}$, where the first contribution is the vacuum term, which reads in the flavor basis:
\begin{equation}
 H_{V}(q) = \frac{1}{2q}\,U_V\,\mathbb{M}^2\, U_V^{\dagger} \, ,
\end{equation}
with $\mathbb{M}$ is the diagonal matrix of neutrino masses, and $U_V$ the unitary vacuum mixing matrix. We have assumed neutrinos to be ultrarelativistic, and discarded a contribution $q \mathbbm{1}$ proportional to the identity in flavor space. For antineutrinos, $U_V$ is replaced by $U_V^*$. In the absence of heavy leptons in the background fluid, the only non-zero element of the second term, the matter Hamiltonian, $H_M$ is $H_M^{(ee)} = \sqrt{2}\,G_F\,n_e$~\cite{1978PhRvD..17.2369W,Mikheyev:1985aa} where $n_e$ is the local electron density. For antineutrinos $\bar{H}_M = - H_M$. Finally, the self-interaction Hamiltonian $H_{SI}$ reads:
\begin{align}
    H_{SI}(t,\vec{r},\vec{q}) &= \sqrt{2} G_F \int{\frac{p^2 dp \, d\Omega_p}{(2 \pi)^3}(1-\widehat{q}\cdot \widehat{p})} \\
    &\qquad \qquad \qquad \times \left[\varrho(t,\vec{r},\vec{p})-\bar{\varrho}^{\star}(t,\vec{r},\vec{p})\right] \nonumber \\
    &= H_E - \mu H_F \, ,
\end{align}
with
\begin{align}
    H_E &= 4 \pi \sqrt{2} G_F \int{\frac{dp}{p}\left[E(t,r,p) - \bar{E}^{\star}(t,r,p)\right]} \, , \\
    H_F &= 4 \pi \sqrt{2} G_F \int{\frac{dp}{p}\left[F_r(t,r,p) - \bar{F}_r^{\star}(t,r,p)\right]} \, .
\end{align}
For antineutrinos, $\bar{H}_{SI} = - H_{SI}^*$.

Integrating Eq.~\eqref{eq:QKE_multi} over the angular direction of $\vec{q}$ and using the definitions \eqref{eq:defE}--\eqref{eq:defP} gives the evolution equations for the first few moments (see also~\cite{2013PhRvD..88j5009Z}): 
\begin{widetext}
\begin{subequations}
\label{eq:QKE_moments}
\begin{align}
\frac{\partial E}{\partial t} + \frac{\partial F_{r}}{\partial r} + \frac{2 F_{r}}{r}  &= -\imath \left[ H_{V}+H_M + H_E,E \right]+\imath \left[ H_F,F_{r}\right] \label{eq:energy} \, , \\
\frac{\partial \bar{E}}{\partial t} + \frac{\partial \bar{F}_{r}}{\partial r} + \frac{2 \bar{F}_{r}}{r}  &= -\imath \left[ H_{V}-H_M-H_{E}^{*},\bar{E} \right] -\imath \left[ H_{F}^{*},\bar{F}_{r}\right] \label{eq:energybar} \, , \\
\frac{\partial F_{r}}{\partial t} + \frac{\partial P_{rr}}{\partial r} + \frac{3\,P_{rr} - E}{r}  &= - \imath \left[H_{V}+H_M+H_{E},F_{r}\right] +\imath \left[H_{F},P_{rr}\right] \label{eq:flux} \, , \\
\frac{\partial \bar{F}_{r}}{\partial t} + \frac{\partial \bar{P}_{rr}}{\partial r} + \frac{3\,\bar{P}_{rr} -\bar{E}}{r}  &= - \imath \left[H_{V}-H_M-H_{E}^{*},\bar{F}_{r}\right] -\imath \left[H_{F}^{*},\bar{P}_{rr}\right] \label{eq:fluxbar} \, .
\end{align}
\end{subequations}
\end{widetext}
As is explicit in Eqs.~\eqref{eq:flux} and \eqref{eq:fluxbar}, the evolution equations for the energy density and flux moments depend upon the pressure moment whose evolution equation is not provided unless we go higher in the tower of moment equations, which would then involve the unevolved rank-3 moment, and so on. In order to solve the equations we need to provide a closure relation that allows determination of the `missing' moment in a truncated tower of evolution equations. There are many different closure schemes found in the literature: for a review of classical closures we refer the reader to Refs.~\cite{Smit_closure_2000,Murchikova:2017zsy}. In a previous paper outlining our code for solving these quantum moment equations in steady-state situations, we considered the very simple choices of $E = K_F\,F$ and $E = K_P\,P$ with $K_F$ and $K_P$ both scalar functions \cite{Myers:2021hnp}. For both these closures, the off-diagonal elements of the related moments, either $E$ and $F$ or $E$ and $P$, are in phase i.e., they are coherent. We found from the test cases we considered that scalar closures could not always reproduce results from a steady-state multi-angle code and conjectured that breaking the phase coherence between the moments would be necessary to improve the agreement. Since the phase breaking cannot be accomplished with a scalar function, we require a more general, quantum, closure between the moments.

\section{Quantum Closures}
\label{sec:quantum_closure}

\subsection{General Parameterization}
\label{subsec:parameterization_closure}

We shall start our derivation of the general properties of a quantum closure by considering the simple case of a relation between the moments $E$ and $P_{rr}$ and the ansatz that they are related via
\begin{equation}
P_{rr} = L\,E\,R \, , \label{eq:ansatz}
\end{equation}
where $L$ and $R$ are two matrices. Since $P_{rr}$ and $E$ are both Hermitian, we require $L\,E\,R = R^{\dagger}\,E\,L^{\dagger}$. It is tempting to look at this requirement and conclude that the relationship between $L$ and $R$ must be $R=L^{\dagger}$, however this is not the only solution. Since $E$ is Hermitian it has an eigenvalue matrix $\Lambda_E$ related to $E$ by a unitary matrix $U_E$, i.e. $E = U_E\,\Lambda_E\,U_E^{\dagger}$. Inserting this in place of $E$ in Eq.~\eqref{eq:ansatz} we realize that a more general solution to the relationship between $L$ and $R$ is
\begin{equation}
R = U_{E}\,D\,U_{E}^{\dagger}\,L^{\dagger} \, ,
\label{eq:R}
\end{equation}
where $D$ is a diagonal matrix of real numbers. The extra term $U_{E}\,D\,U_{E}^{\dagger}$ in equation (\ref{eq:R}) is allowed because it commutes with $U_E\,\Lambda_E\,U_E^{\dagger}$. Although there is no unique relation between $L$ and $R$, setting $D$ to be the unit matrix $\mathbbm{1}$ so that $R=L^{\dagger}$ is the most convenient relation for now and so we define the quantum closure in the form:
\begin{equation}
P_{rr} = L\,E\,L^{\dagger} \, ,
\label{eq:quantumclosure}
\end{equation}
which is the statement that the two moments used in the closure are Hermitely congruent, or conjunctive. At the end of this section we shall show how $D \neq \mathbbm{1}$ generally emerges from a quantum closure and indeed, there are circumstances where one would need to make use of ${D \neq \mathbbm{1}}$. 

Our quest for a quantum closure now turns to finding an appropriate form for $L$. This can be done if we assume we know $E$ and $P_{rr}$ and have to compute a matrix which relates them according to Eq.~\eqref{eq:quantumclosure}. First, note that $E$ and $P_{rr}$ are both positive-semidefinite Hermitian matrices. \newtext{Indeed, recall that the one-body density matrix $\varrho(t,\vec{r},\vec{q})$ is, by definition, a positive-semidefinite operator~\cite{cohen2019quantum}. Any (discrete or continuous) linear combination of such operators, with positive coefficients, is still positive-semidefinite. This is the case, in particular, for $E$ and $P_{rr}$ [see Eqs.~\eqref{eq:defE} and \eqref{eq:defP}]. This proves that these two moments are also positive-semidefinite — note that the same cannot be said for $F_r$, since the coefficient multiplying $\varrho$ in \eqref{eq:defF} takes both positive and negative values.} 
We may then write $E$ and $P_{rr}$ as $E=\epsilon\,\epsilon^{\dagger}$ and $P_{rr}=\rho\,\rho^{\dagger}$. A Cholesky decomposition is an example of this matrix factorization but, in general, the factorization is not unique unless we impose additional constraints. Nevertheless, using such a decomposition, the solution for $L$ is
\begin{equation}
L = \rho\,\epsilon^{-1} \, . 
\label{eq:L1}
\end{equation}
The matrix $\epsilon$ can be identified from the requirement that $E = \epsilon\,\epsilon^{\dagger} = U_E\, \Lambda_E\, U_E^{\dagger}$. The general solution for $\epsilon$ is therefore $\epsilon = U_E\,\Lambda^{1/2}_E\,S_E$ with $\Lambda^{1/2}_E$ the diagonal matrix whose elements are the positive square roots of the diagonal elements of $\Lambda_E$ i.e.~the matrix square root of $\Lambda_E$, and $S_E$ an arbitrary unitary matrix. Similarly, if $\Lambda_P$ is the matrix of the eigenvalues of $P_{rr}$ and $U_P$ the unitary matrix such that $P_{rr} = U_P\, \Lambda_P\, U_P^{\dagger}$, then we write $\rho = U_P\,\Lambda^{1/2}_P\,S_P$. Inserting these relations into Eq.~\eqref{eq:L1} we find
\begin{equation}
L = U_P\,\Lambda^{1/2}_P \,\,S_P \,S_E^{\dagger}\,\Lambda^{-1/2}_E \, U_E^{\dagger} \, .
\end{equation}
The matrices $U_E$ and $U_P$ can be factored as $U_E = \Upsilon_E\,\Psi_E$ and $U_P=\Upsilon_P\,\Psi_P$ where $\Psi_E$ and $\Psi_P$ are diagonal matrices of complex exponentials with arbitrary (Majorana) phases. The matrices $\Upsilon_P$ and $\Upsilon_E$ contain all the parts of $U_P$ and $U_E$ that can be uniquely determined by the requirement that they diagonalize $P_{rr}$ and $E$ respectively \emph{and} that each column of $U_P$ and/or $U_E$ — which are the eigenvectors $E$ and $P_{rr}$ — satisfy an additional phase constraint e.g. that the entry in the first top row is positive and real. 
Inserting this factorization, using the fact that $\Psi_E$ and $\Psi_P$ commute with $\Lambda_{P}^{1/2}$ and $\Lambda_{E}^{-1/2}$ since all are diagonal, and choosing the arbitrary matrices $S_E$ and $S_P$ so that $\Psi_P\,S_P\,S^{\dagger}_E\,\Psi_E^{\dagger} =  \mathbbm{1}$, we find
\begin{equation} 
L = \Upsilon_P\,X^{1/2}\;\Upsilon^{\dagger}_E \, ,
\label{eq:L}
\end{equation}
where we have introduced the matrix $X^{1/2} = \Lambda_P^{1/2} \Lambda_E^{-1/2}$. 
Inserting this result into Eq.~\eqref{eq:quantumclosure} for the closure relation and gathering terms as 
\begin{equation}
P_{rr} = \Upsilon_P \left( X^{1/2}\, \Bigl( \Upsilon^{\dagger}_E\,E\,\Upsilon_E \Bigr)\, X^{1/2} \right) \Upsilon^{\dagger}_P \, ,
\label{eq:QC}
\end{equation}
we observe the general structure of the quantum closure relationship: the innermost term $\Upsilon^{\dagger}_E\,E\,\Upsilon_E$ produces the eigenvalue matrix $\Lambda_E$; the term $X^{1/2}\,\Lambda_E\,X^{1/2}$ rescales the eigenvalue matrix $\Lambda_E$ to the eigenvalue matrix $\Lambda_P$, and $\Upsilon_P\,\Lambda_P\,\Upsilon^{\dagger}_P$ `de-diagonalizes' $\Lambda_P$. 
While $\Upsilon_E$ appears in Eq.~\eqref{eq:L} for $L$, $\Upsilon_E$ can always be computed given the moment $E$ so it is not independent of the moment and thus should not be regarded as introducing new parameters into the closure. Thus we find a quantum closure is a function of $N^2_f$ parameters where $N_f$ is the number of neutrino flavors: the matrix $X$ has $N_f$ parameters, and $\Upsilon_P$ is a function of $N_f^2 - N_f$ parameters. 

In the case of two neutrino flavors the two matrices $\Upsilon_E$ and $\Upsilon_P$ can be easily expressed in terms of the elements of $E$ and $P_{rr}$. As shown in Appendix~\ref{sec:app1}, both $E$ and $P_{rr}$ can be expanded in terms of a set of Hermitian basis matrices $\{ e^{\,\widetilde{t}},e^{\,\widetilde{x}},e^{\,\widetilde{y}},e^{\,\widetilde{z}}\}$ with real coefficients:
\begin{equation}
\label{eq:M_decompos}
M = M^{\,\widetilde{t}}\,e^{\,\widetilde{t}} + M^{\,\widetilde{x}}\,e^{\,\widetilde{x}} + M^{\,\widetilde{y}}\,e^{\,\widetilde{y}} + M^{\,\widetilde{z}}\,e^{\,\widetilde{z}} \, ,
\end{equation}
with $M \in \{ E, P_{rr} \}$. 
We also introduce for later convenience the hyperspherical angles $\eta_M$, $\theta_M$, $\phi_M$: 
\begin{equation}
\begin{aligned}
M^{\,\widetilde{t}} &= ||M||_F\,\cos\eta_M \\
M^{\,\widetilde{x}} &= ||M||_F\,\sin\eta_M\,\sin\theta_M\,\cos\phi_M\\
M^{\,\widetilde{y}} &=  ||M||_F\,\sin\eta_M\,\sin\theta_M\,\sin\phi_M\\
M^{\,\widetilde{z}} &= ||M||_F\,\sin\eta_M\,\cos\theta_M ,
\label{eq:Mcomponents}
\end{aligned}
\end{equation}
where $||M||_F$ is the Frobenius norm of $M$ equal to 
$||M||_F^2 = (M^{\,\widetilde{t}})^2 + (M^{\,\widetilde{x}})^2 + (M^{\,\widetilde{y}})^2 + (M^{\,\widetilde{z}})^2$. The eigenvalue matrix for $M$ is 
\begin{align}
\Lambda_M &=  M^{\,\widetilde{t}}\,e^{\,\widetilde{t}} + |\vec{M}|\,e^{\,\widetilde{z}}  \nonumber \\ &=  M^{\,\widetilde{t}}\,\left( e^{\,\widetilde{t}} + v_M\,e^{\,\widetilde{z}} \right) \, ,
\end{align}
where $\vec{M}$ is the 3-vector formed by the `space' components of $M$ i.e. $\vec{M} = (M^{\,\widetilde{x}}, M^{\,\widetilde{y}},M^{\,\widetilde{z}})$ and the ‘speed’ $v_M = \lvert \vec{M} \rvert/M^{\, \widetilde{t}} = \lvert \tan(\eta_M) \rvert$.

Following the procedure for computing $U$ given in \cite{2012JPhG...39c5201G} and using the notation introduced above, we find

\begin{widetext}
\begin{align}
U_E &= \frac{1}{ \sqrt{2\,v_E\,E^{\,\widetilde{t}}\,\left(v_E\,E^{\,\widetilde{t}}+E^{\,\widetilde{z}}\right) }} 
\begin{pmatrix} v_E\,E^{\,\widetilde{t}}+E^{\,\widetilde{z}} & E^{\,\widetilde{x}}-\imath\,E^{\,\widetilde{y}}   \\ E^{\,\widetilde{x}}+\imath\,E^{\,\widetilde{y}} & -v_E\,E^{\,\widetilde{t}}-E^{\,\widetilde{z}}   \end{pmatrix} \,\Psi_E \, = \Upsilon_E\,\Psi_E , \\
U_P &= \frac{1}{ \sqrt{2\,v_P\,P_{rr}^{\,\widetilde{t}}\,\left(v_P\,P_{rr}^{\,\widetilde{t}}+P_{rr}^{\,\widetilde{z}}\right) }}
\begin{pmatrix} v_P\,P_{rr}^{\,\widetilde{t}}+P_{rr}^{\,\widetilde{z}} & P_{rr}^{\,\widetilde{x}}-\imath\,P_{rr}^{\,\widetilde{y}}   \\ P_{rr}^{\,\widetilde{x}}+\imath\,P_{rr}^{\,\widetilde{y}} & -v_P\,P_{rr}^{\,\widetilde{t}}-P_{rr}^{\,\widetilde{z}}   \end{pmatrix} \, \Psi_P = \Upsilon_P\,\Psi_P\,  ,
\end{align}
\end{widetext}
where $v_E = |\vec{v}_E| = | \vec{E} / E^{\,\widetilde{t}}|$ and similarly for $v_P$, and $\Psi_E$ and $\Psi_P$ are the diagonal matrices of arbitrary phases. Thus we are able to identify the other terms in these equations as $\Upsilon_E$ and $\Upsilon_P$ and observe they are traceless, Hermitian and unitary matrices.\footnote{The traceless and Hermitian properties are unique to $N_f =2$.} As shown in Appendix~\ref{sec:app1}, they are functions of $\theta_E,\phi_E$ and $\theta_P,\phi_P$ only. Thus for $N_f =2$ flavors, the quantum closure is seen to be a function of four parameters: the two diagonal elements of $X$, and the two parameters $\theta_P$ and $\phi_P$ needed to define $\Upsilon_P$. Though the parameters needed to describe $\Upsilon_E$ are already known if $E$ is provided, we nevertheless include them in the closure because they serve as useful reference values. 

There are two ways to parameterize $X$: first we can define 
\begin{align}
\label{eq:Xmatrix_12}
X &\equiv \begin{pmatrix} \chi_1 & 0  \\  0 & \chi_2 \end{pmatrix} \, , \\ \nonumber \\ 
\intertext{and after writing $L$ as} 
L &= \begin{pmatrix} L_{11} & L_{12}  \\  L_{21} & L_{22} \end{pmatrix} \, ,
\end{align}
we obtain

\begin{widetext}
\begin{subequations}
\label{eq:closure_L}
\begin{align}
L_{11} &= \sqrt{\chi_1}\,\Upsilon_E^{\,\widetilde{z}}\,\Upsilon_P^{\,\widetilde{z}} + \sqrt{\chi_2}\, \left(\Upsilon_E^{\,\widetilde{x}} + \imath \Upsilon_E^{\,\widetilde{y}}\right)\,\left(\Upsilon_P^{\,\widetilde{x}} - \imath \Upsilon_P^{\,\widetilde{y}} \right) \label{eq:L11} \\
  &=  \sqrt{\chi_1}\,\cos\left(\frac{\theta_E}{2}\right)\,\cos\left(\frac{\theta_P}{2}\right) + \sqrt{\chi_2}\,\sin\left(\frac{\theta_E}{2}\right)\,\sin\left(\frac{\theta_P}{2}\right)\,e^{\imath(\phi_E-\phi_P)} \\
L_{12} &= \sqrt{\chi_1}\,\Upsilon_P^{\,\widetilde{z}}\,\left(\Upsilon_E^{\,\widetilde{x}} - \imath \Upsilon_E^{\,\widetilde{y}}\right) - \sqrt{\chi_2}\, \Upsilon_E^{\,\widetilde{z}}\,\left(\Upsilon_P^{\,\widetilde{x}} - \imath \Upsilon_P^{\,\widetilde{y}}\right) \\
 &=  \sqrt{\chi_1}\,\sin\left(\frac{\theta_E}{2}\right)\,\cos\left(\frac{\theta_P}{2}\right)\,e^{-\imath\phi_E} - \sqrt{\chi_2}\,\cos\left(\frac{\theta_E}{2}\right)\,\sin\left(\frac{\theta_P}{2}\right)\,e^{-\imath\phi_P} \\
L_{21} &=  \sqrt{\chi_1}\,\Upsilon_E^{\,\widetilde{z}}\,\left(\Upsilon_P^{\,\widetilde{x}} + \imath \Upsilon_P^{\,\widetilde{y}}\right) - \sqrt{\chi_2} \,\Upsilon_P^{\,\widetilde{z}}\,\left(\Upsilon_E^{\,\widetilde{x}} + \imath \Upsilon_E^{\,\widetilde{y}}\right) \\
&= \sqrt{\chi_1}\,\cos\left(\frac{\theta_E}{2}\right)\,\sin\left(\frac{\theta_P}{2}\right)\,e^{\imath\phi_P} - \sqrt{\chi_2}\,\sin\left(\frac{\theta_E}{2}\right)\,\cos\left(\frac{\theta_P}{2}\right)\,e^{\imath\phi_E}  \\
L_{22} &= \sqrt{\chi_1}\, \left(\Upsilon_E^{\,\widetilde{x}} - \imath \Upsilon_E^{\,\widetilde{y}}\right)\,\left(\Upsilon_P^{\,\widetilde{x}} + \imath \Upsilon_P^{\,\widetilde{y}} \right)  + \sqrt{\chi_2}\,\,\Upsilon_E^{\,\widetilde{z}} \Upsilon_P^{\,\widetilde{z}} \\ 
&= \sqrt{\chi_1}\,\sin\left(\frac{\theta_E}{2}\right)\,\sin\left(\frac{\theta_P}{2}\right)\,e^{-\imath(\phi_E-\phi_P)} + \sqrt{\chi_2}\,\cos\left(\frac{\theta_E}{2}\right)\,\cos\left(\frac{\theta_P}{2}\right) \label{eq:L22}. 
\end{align}
\end{subequations}
\end{widetext}
In the second line of each pair of equations above we have inserted the expressions for the components of $\Upsilon_E$ and $\Upsilon_P$ from Appendix~\ref{sec:app1}. Alternatively we can write $X$ as 
\begin{equation}
\label{eq:Xmatrix_chiv}
X = \chi \begin{pmatrix} \dfrac{1+v_P}{1+v_E} & 0  \\  0 & \dfrac{1-v_P}{1-v_E} \end{pmatrix} \, ,
\end{equation}
where $\chi$ and $v_P$ are the two free parameters and $v_E$ is computed from the moment $E$. 
The connection between the two parameterizations of the $X$ matrix, \eqref{eq:Xmatrix_12} and \eqref{eq:Xmatrix_chiv}, reads:
\begin{equation}
    \begin{aligned}
        \chi &= \frac12 \left[\chi_1(1+v_E)+\chi_2(1-v_E)\right] \, , \\
        \chi \, v_P &= \frac12 \left[\chi_1(1+v_E) - \chi_2(1-v_E)\right] \, .
    \end{aligned}
\end{equation} 
In this alternative form the parameter $\chi = P^{\,\widetilde{t}} / E^{\,\widetilde{t}}$ is a `classical' parameter in the sense that neutrino oscillations do not alter this term, but it differs from the usual meaning of an Eddington ratio because it includes both electron neutrinos \emph{and} the $x$ flavor. This second parameterization is useful because, when inserted into equations~\eqref{eq:L11} through \eqref{eq:L22}, and then express $E$ and $P_{rr}$ in terms of their components~\eqref{eq:M_decompos}, we find the succinct expressions:
\begin{subequations}
\label{eq:components_P}
\begin{align}
P_{rr}^{\,\widetilde{t}} &= \chi\,E^{\,\widetilde{t}} \, , \label{eq:Pt_Et} \\
P_{rr}^{\,\widetilde{x}} &= v_P\,\chi\,E^{\,\widetilde{t}}\,\sin\theta_P\,\cos\phi_P  \, , \\
P_{rr}^{\,\widetilde{y}} &= v_P\,\chi\,E^{\,\widetilde{t}}\,\sin\theta_P\,\sin\phi_P  \, , \\
P_{rr}^{\,\widetilde{z}} &= v_P\,\chi\,E^{\,\widetilde{t}}\,\cos\theta_P \, .
\end{align}
\end{subequations}

Before proceeding, we give some final remarks about quantum closures in general, not just the $N_f=2$ case. First, as stated at the beginning of this section, there appears to be an additional degree of freedom in the closure that we have ignored when we set the matrix $D$ in Eq.~\eqref{eq:R} to the unit matrix. This inference is not correct although there is certainly an ambiguity in a quantum closure as we now show. We observe that given the decomposition of $L$ in Eq.~\eqref{eq:L} and that $X^{1/2}$ and $\Upsilon_E^\dagger\,E\,\Upsilon_E$ commute, one could instead write Eq.~\eqref{eq:QC} in two different ways: 
\begin{equation}
P_{rr} = \Upsilon_P\,X\,\Upsilon^{\dagger}_E\,E\,\Upsilon_E\, \Upsilon^{\dagger}_P \, ,
\end{equation}
or 
\begin{equation}
P_{rr} = \Upsilon_P\,\Upsilon^{\dagger}_E\,E\,\Upsilon_E\,X\,\Upsilon^{\dagger}_P \, .
\label{eq:altQC}
\end{equation}
This shows that we could have assigned instead $L = \Upsilon_P\,\Upsilon^{\dagger}_E$ and $R = \Upsilon_E\,X\,\Upsilon^{\dagger}_P = \Upsilon_E\,X\,\Upsilon^{\dagger}_E\, L^{\dagger}$. This assignment for $R$ is exactly of the general form (i.e.~$D$ not equal to the unit matrix) given in Eq.~\eqref{eq:R}. Thus there is not really an additional degree of freedom missing from Eq.~\eqref{eq:QC} but rather there is an ambiguity arising from the decision about which order to write the terms in the equation. 

Secondly, our derivation of a general quantum closure considered only a relationship between the moments $E$ and $P_{rr}$ but we also note in passing that it is also possible to consider a relationship between $P_{rr}$ and $F_r$, or a combination of the two (this is notably what appears naturally in the extension of the quantum maximum entropy closure for small flavor coherence, see~\cite{Froustey_QMinerbo}). The issue we encounter when using $F_r$ instead of $E$ is that $F_r$ does not have to be a positive-semidefinite matrix, hence its eigenvalues are not guaranteed to be positive. Therefore, the square root of its eigenvalue matrix, $\Lambda_F$, would not be Hermitian. Thus the form for $L$ given in Eq.~\eqref{eq:L} cannot be used but, appealing to Eq.~\eqref{eq:altQC}, a viable form for $L$ and the relationship between $P_{rr}$ and $F_r$, valid even if the eigenvalues of $F_r$ are not positive, is
\begin{equation}
P_{rr} = \Upsilon_P\,Y\,\Upsilon^{\dagger}_F\,F_r\,\Upsilon_F\, \Upsilon^{\dagger}_P =  \Upsilon_P\,\Upsilon^{\dagger}_F\,F_r\,\Upsilon_F\,Y\,\Upsilon^{\dagger}_P.
\end{equation}
where $\Upsilon_F$ is the reduced unitary matrix that diagonalizes $F_r$ to its eigenvalue matrix $\Lambda_F$, and we define $Y = \Lambda_P\,\Lambda_F^{-1}$.


\subsection{Limits on the Closure Parameters}
\label{subsec:limits_params}

For a closure relating the energy density $E$ and the pressure $P_{rr}$, the parameters are limited to ranges in order that the moments be physically realizable. One set of limits arises because both $P_{rr}$ and $E$ are positive-semidefinite matrices; another set arises because $E$ and $P_{rr}$ are related independently of the closure.  

The positive-semidefinite nature of $P_{rr}$ and $E$ can be expressed as
\begin{eqnarray}
    \begin{aligned}
    0 &\leq P_{rr}^{\,\widetilde{t}},\\
    0 &\leq |\vec{P}_{rr}| \leq P_{rr}^{\,\widetilde{t}},
    \end{aligned}
    \label{eq:nevernegative}
\end{eqnarray}
and similarly for elements of the energy density. This is more intuitively true in classical flavor-diagonal distributions, and this requirement is more generally equivalent to stating that the diagonal components of each tensor must be positive in \textit{any} basis. These constraints translate into the requirement that $\chi \geq 0$, $0 \leq v_P \leq 1$. 
The Frobenius norm of $P_{rr}$ is $||P_{rr}||_F = P_{rr}^{\,\widetilde{t}}\,\sqrt{1+v_P^2}$ so we see that the norm lies in the range $P_{rr}^{\,\widetilde{t}} \leq ||P_{rr}||_F \leq \sqrt{2} \,P_{rr}^{\,\widetilde{t}}$. Thus the angle $\eta_P$ must lie in the range $0 \leq \eta_P \leq \pi/4$, and similarly for $\eta_E$. 

Additional limits arise because $E$ and $P_{rr}$ are related. By construction, the spatial trace of any moment reduces the rank of the moment by two. That is, 
\begin{equation}
    \sum_{i} P_{ii} = E \, .
    \label{eq:trace}
\end{equation}
Thus, using spherical coordinates and assuming axial symmetry around the radial direction, we find
\begin{equation}
E = P_{rr}+P_{\theta\theta}+P_{\phi\phi} = P_{rr}+2\,P_{\theta\theta} \, ,
\end{equation}
which applies to each flavor component of the moments. Since $P_{\theta\theta}$ is also positive-semidefinite, i.e. $0 \leq P_{\theta\theta}^{\,\widetilde{t}}$ and $|\vec{P}_{\theta\theta}| \leq P_{\theta\theta}^{\,\widetilde{t}}$, this immediately tells us that the closure must be such that $P_{rr}^{\,\widetilde{t}} \leq E^{\,\widetilde{t}}$ and that $|\vec{E} - \vec{P}_{rr}| \leq E^{\,\widetilde{t}} - P_{rr}^{\,\widetilde{t}}$. The first constraint reads $\chi \leq 1$. The second constraint leads to limits on the angles $\theta_P$ and $\phi_P$ which requires a short digression into matrix alignment. 

Independent of the closure, the degree of similarity between the two moments $E$ and $P_{rr}$ can be measured by the degree to which they `align'. The notions of matrix alignment can be defined in various ways. As shown in Appendix~\ref{sec:app1}, we can define the `angle' $\Xi$ between the energy density $E$ and the pressure tensor component $P_{rr}$ to be
\begin{align}
    \label{eq:Xi} 
\cos \Xi &\equiv \frac{\langle E, P_{rr} \rangle_{F} }{||E||_F \, ||P_{rr}||_F} \\
   &= \cos\eta_E\,\cos\eta_P + \sin\eta_E\,\sin\eta_P \, \times  \nonumber \\ 
   & \quad \left[ \cos\theta_E\,\cos\theta_P + \sin\theta_E\,\sin\theta_P\,\cos(\phi_E-\phi_P)\right] \, . \nonumber  
\end{align}
However this is not the only alignment angle one can define. 
Since one can isolate the `spatial' part of a moment in flavor space by subtracting the `time' component, we can also define an angle $\xi$ between the spatial parts of $E$ and $P_{rr}$ as
\begin{align}
\label{eq:cosxi}
\cos \xi &\equiv \frac{\langle E -E^{\,\widetilde{t}}\,e^{\,\widetilde{t}}, P_{rr} -P_{rr}^{\,\widetilde{t}}\,e^{\,\widetilde{t}} \rangle_{F} }{||E-E^{\,\widetilde{t}}\,e^{\,\widetilde{t}}||_F \, ||P_{rr}-P_{rr}^{\,\widetilde{t}}\,e^{\,\widetilde{t}}||_F} \nonumber \\ 
& = \frac{\vec{E}\cdot\vec{P}_{rr}}{|\vec{E}|\,|\vec{P}_{rr}|} \\
&= \cos\theta_E\,\cos\theta_P + \sin\theta_E\,\sin\theta_P\,\cos\left(\phi_E-\phi_P\right) \, . \nonumber
\end{align}
Thus $\Xi$ and $\xi$ are related via 
\begin{equation}
\cos \Xi = \cos\eta_E\,\cos\eta_P + \sin\eta_E\,\sin\eta_P\,\cos\xi \, .
\label{eq:xirelation}
\end{equation}

Returning to the constraint $|\vec{E} - \vec{P}_{rr}| \leq E^{\,\widetilde{t}} - P_{rr}^{\,\widetilde{t}}$, we find this is equivalent to the requirement that the alignment angle $\Xi$ satisfy
\begin{equation}
\cos \Xi \geq \frac{ 2\,\chi + v_E^2 +  \chi^2\,v_P^2 - (1-\chi)^2}{ 2\,\chi\,\sqrt{1+v_E^2}\,\sqrt{1+v_P^2} } \, ,
\label{eq:cosXilimit}
\end{equation}
and / or that $\xi$ must satisfy
\begin{equation}
\cos \xi \geq \frac{v_E^2 + \chi^2 v_P^2 - (1-\chi)^2}{2 \chi v_E v_P} \, .
\label{eq:cosxilimit0}
\end{equation}
The two requirements can be shown to be related. From Eq.~\eqref{eq:cosXilimit}, one may show that for $1 > \chi > 1/3$ the minimum value of the right hand side occurs at $v_E=0$ and $v_P = \sqrt{(1-3\,\chi)(\chi-1)}/\chi$, so that
\begin{equation}
\cos \Xi \geq \sqrt{4\,\chi-2\,\chi^2-1} \, .
\label{cosXilimitlargechi}
\end{equation}
For $\chi<1/3$, the minimum is at $v_E = v_P=0$, and we obtain
\begin{equation}
\cos \Xi \geq \frac{4\,\chi-\chi^2-1}{2\,\chi} \, .
\label{cosXilimitsmallchi}
\end{equation}
Using the relation between $\Xi$ and $\xi$ we can turn this result into  
\begin{eqnarray}
\label{eq:cosxilimit2}
\cos \xi & \geq & \frac{\sqrt{1+v_E^2}\,\sqrt{1+v_P^2}\,\cos\Xi_{\mathrm{max}} - 1}{v_E\,v_P} \, ,
\end{eqnarray}
where $\cos\Xi_\mathrm{max}$ is the right hand side of either Eq.~\eqref{cosXilimitlargechi} or \eqref{cosXilimitsmallchi} given the value of $\chi$. 
Both equations \eqref{eq:cosxilimit0} and \eqref{eq:cosxilimit2} show that the angle between spatial parts of $E$ and $P_{rr}$ is limited for given values of $\chi$ and $v_P$. If $\xi_\mathrm{max}$ is the largest value of $\xi$ that satisfies Eq.~\eqref{eq:cosxilimit2}, then from Eq.~\eqref{eq:cosxi} we see that $\theta_P$ and $\phi_P$ must be such that 
\begin{equation}
\cos\theta_E\,\cos\theta_P + \sin\theta_E\,\sin\theta_P\,\cos\left(\phi_E-\phi_P\right) 
\geq \cos\xi_\mathrm{max} \, .  
\end{equation}
From this result we find $|\theta_E - \theta_P| \leq \xi_\mathrm{max}$ and the difference $\phi_E-\phi_P$ must satisfy
\begin{eqnarray}
\cos\left(\phi_E-\phi_P\right) 
& \geq & \frac{ \cos\xi_\mathrm{max} - \cos^2\theta_E }{\sin^2\theta_E}.  
\end{eqnarray}
At $\theta_E = \pi/2$ the absolute value of the difference between the two phase angles is restricted to be $|\phi_E-\phi_P| \leq \xi_\mathrm{max}$ but at other values of $\theta_E$, the difference $|\phi_E-\phi_P|$ can be larger than $\xi_\mathrm{max}$.  

\paragraph*{Limiting regimes ---} We end this section by commenting on the two physical limits of optically thick and thin regimes. In the diffusive limit, for which distributions are isotropic, Eq.~\eqref{eq:defP} reads $P_{rr}^{(ab)} = E^{(ab)}/3$ for all pairs $(a,b)$ of flavors, which corresponds to $\chi=1/3$, $v_P=v_E$ (so $\eta_P = \eta_E$), $\theta_P = \theta_E$ and $\phi_P = \phi_E$. The alignment angles~\eqref{eq:Xi} and \eqref{eq:cosxi} are then $\cos \xi = \cos \Xi = 1$. In the free-streaming regime, we have $P_{rr}^{(ab)} = E^{(ab)}$ which corresponds to $\chi=1$, $v_P=v_E$, $\theta_P = \theta_E$ and $\phi_P = \phi_E$. Once again, $\cos \xi = \cos \Xi = 1$. The richness of possible quantum closures, encoded in our parameterization, then reveals itself in the intermediate regimes.


\section{Numerical Calculations}
\label{sec:numerical_calculations}

With our general framework in hand, we continue to explore quantum closures by undertaking several numerical calculations. The first goal of these calculations is to derive the quantum closure from approaches which allow us to determine the moments independently i.e. we are able to find the distribution matrices $\varrho$ and $\bar{\varrho}$ at a given location and compute the moments, and our second goal is to try out the quantum closures we find in a moment based evolution code to examine if the use of quantum closures improves the agreement between moment based evolution and the other methods. 

The first goal is the challenge. Codes which allow us to compute in non-trivial scenarios the density matrices $\varrho$ and $\bar{\varrho}$ - what we shall call multi-angle codes - are naturally computationally expensive. Scenarios which involve time \emph{and} spatial variations of the moments introduce additional challenges associated with advection that we wish to avoid. For these reasons, we consider two cases where the evolution equations for the multi-angle problem and the moments reduce to ODEs. If we consider a steady-state then we remove the time dependence and the simplest non-trivial steady-state multi-angle scenario is the Bulb Model first solved by Duan \emph{et al.}~\cite{2006PhRvL..97x1101D,Duan:2006an}. Alternatively, if we assume homogeneity then we remove the spatial dependence and the evolution equations reduce to ODEs in time.


\subsection{Steady-State MSW Problem}
\label{sec:MSWproblem}

We start by considering the steady-state multi-angle Bulb Model problem where only the vacuum and matter terms contribute to the Hamiltonian $H$ — see Sec.~\ref{sec:EvolutionEquations} for the expressions. This same test problem was considered previously in \cite{Myers:2021hnp}. We re-consider it here because a) it has an analytic solution \cite{Blennow:2013rca} that allows us to verify our codes, and b) the multi-angle and moment calculations do not agree when a scalar closure is used, indicating the necessity of using a quantum closure. 
The geometry of the Bulb Model is well known but we include it here for completeness. The neutrino and antineutrino emission occurs from a spherical neutrinosphere with a radius $R_{\nu}$. Neutrinos which are emitted at an angle $\theta_R$ relative to the radial direction at the emission surface make an angle $\theta$ with respect to the radial direction at the radial location $r$ above the source. In a flat spacetime the angles $\theta$ and $\theta_R$ are related via 
\begin{equation}
\cos\theta = \sqrt{1- \left(\frac{R_{\nu}\,\sin\theta_R}{r}\right)^2}.
\end{equation}
Also in a flat spacetime, the neutrinos have traveled a distance $\lambda$ given by  
\begin{equation}
\lambda = r\cos\theta - \sqrt{R_{\nu}^2 - r^2\,\sin^2 \theta}
\end{equation}
from their emission point and if they were emitted with energy $q_R$ then their energy at radius $r$ will be the same i.e. $q=q_R$. The flavor evolution of neutrinos in a curved spacetime has been considered elsewhere~\cite{1996GReGr..28.1161A,1996PhRvD..54.1587P,1997PhRvD..55.7960C,1997PhRvD..56.1895F,1998PThPh.100.1145K,2005PhRvD..71g3011L,2013JCAP...06..015D,2015GReGr..47...62V,2016NuPhB.911..563Z,2017PhRvD..96b3009Y,2020PhLB..80135150C,2021GReGr..53...98S,2022PhRvD.106f3011N}, but for now we relegate quantum closures in curved spacetime to future considerations. 

\subsubsection{Numerical setup}

The density matrices $\varrho(R_{\nu},\theta_R,q_R)$ and $\bar{\varrho}(R_{\nu},\,\theta_R,q_R)$ for the neutrinos and antineutrinos at the emission surface are taken to be diagonal in flavor space, consistently with the emission of (anti)neutrinos in pure flavor states. The diagonal elements of the density matrices are typically taken to be of the form 
\begin{equation}
\label{eq:init_distrib}
\varrho^{(aa)}(R_{\nu},\theta_R,q_R) = A^{(a)}\, Q^{(a)}(q_R)\,\Theta^{(a)}(\theta_R) \, ,
\end{equation}
where $A^{(a)}$ is a normalization constant allowing to get the desired initial flux or luminosity, 
and similarly for the antineutrinos. $Q^{(a)}$ is a function describing the energy spectrum for flavor $a$ and in the numerical test case we shall consider only a single energy so that $Q^{(a)}$ is a delta-function at the chosen energy. $\Theta^{(a)}$ is a function describing the angular spectrum for flavor $a$ and the form of the initial angular distributions is often taken to be that introduced by~\cite{Mirizzi:2011tu},
\begin{equation}
\label{eq:angular_distribution}
\Theta^{(a)}(\theta_R) = (\beta^{(a)} + 2)\, \left[\cos(\theta_R)\right]^{\beta^{(a)}} \, .
\end{equation}
The case of half-isotropic emission corresponds to $\beta^{(a)} = 0$. The factor of $(\beta^{(a)} + 2)$ in Eq.~\eqref{eq:angular_distribution} normalizes the angular distribution. 
For every emission angle (and each energy if we consider a multi-energy calculation), we introduce an evolution matrix $S(r,\theta,q;R_{\nu},\theta_R,q_R)$ that relates the density matrix $\varrho(r,\theta,q)$ at radial coordinate $r$ for neutrinos with energy $q$ propagating at angle $\theta$ relative to the radial direction, to the initial distribution matrices~\eqref{eq:init_distrib}. 
The matrix $\bar{S}(r,\theta,q;R_{\nu},\theta_R,q_R)$ serves in the same role for antineutrinos. The matrix $S(r,\theta,q;R_{\nu},\theta_R,q_R)$ evolves according to the Schr\"odinger equation
\begin{equation}
\imath \frac{dS}{d\lambda} = H\, S \, ,
\end{equation}
where $H$ is the Hamiltonian from Eq.~\eqref{eq:QKE_multi} but we only retain the vacuum and matter terms, and $\lambda$ is the affine parameter along the trajectory. The antineutrinos evolve according to the Hamiltonian $\bar{H}$. 
$S$ and $\bar{S}$ are initially the identity matrices and this does not change if we start our calculation at a radius other than the neutrinosphere and assume no flavor evolution has occurred up to that point.

The evolution equations for the moments were previously given in Eqs.~\eqref{eq:energy} through \eqref{eq:fluxbar} and in the steady-state, these equations become ODEs for the radial component of the differential energy flux vectors $F_r$ and ${\bar{F}}_r$,  and the `$rr$' components of the pressure tensors $P_{rr}$ and $\bar{P}_{rr}$. Thus, in steady-state problems the `missing' moments are actually the energy densities $E$ and $\bar{E}$. If we start our moment calculation at a radius $r_0$ other than the neutrinosphere $R_{\nu}$ and assume no flavor evolution has occurred up that point, the initial conditions for the moments are \cite{Myers:2021hnp}
\begin{widetext}
\begin{align}
E(r_0,q) &= \frac{R_{\nu}^2}{2\,r_0^2}\,\left(\frac{q}{2 \pi}\right)^{3}
\begin{pmatrix} A^{(e)}\,Q^{(e)}(q)\,_{2}F_{1}\left(1,\frac{1}{2},2+\frac{\beta^{(e)}}{2},\frac{R_{\nu}^2}{r_0^2}\right) & 0 \\ 0 & A^{(x)}\,Q^{(x)}(q)\,_{2}F_{1}\left(1,\frac{1}{2},2+\frac{\beta^{(x)}}{2},\frac{R_{\nu}^2}{r_0^2}\right)
    \end{pmatrix} \, , \\
F_r(r_0,q) & = \frac{R_{\nu}^2}{2\,r_0^2}\,\left(\frac{q}{2 \pi}\right)^{3}
\begin{pmatrix} A^{(e)}\,Q^{(e)}(q) & 0 \\ 0 & A^{(x)}\,Q^{(x)}(q)
    \end{pmatrix} \, , \\
P_{rr}(r_0,q) & = \frac{R_{\nu}^2}{2\,r_0^2}\,\left(\frac{q}{2 \pi}\right)^{3}
\begin{pmatrix} A^{(e)}\,Q^{(e)}(q)\,_{2}F_{1}\left(1,-\frac{1}{2},2+\frac{\beta^{(e)}}{2},\frac{R_{\nu}^2}{r_0^2}\right) & 0 \\ 0 & A^{(x)}\,Q^{(x)}(q)\,_{2}F_{1}\left(1,-\frac{1}{2},2+\frac{\beta^{(x)}}{2},\frac{R_{\nu}^2}{r_0^2}\right)
    \end{pmatrix}  \, ,
\end{align}
\end{widetext}
where $_2F_1(a,b,c,z)$ is the ordinary hypergeometric function.

Finally, for this test problem we set the matter density to be a constant of $\rho = 8 \times 10^{3} \, \mathrm{g\,cm^{-3}}$ and electron fraction $Y_e = 0.5$. For the difference of the neutrino squared mass we use $\Delta m^2 = m^{2}_{2} - m^{2}_{1} = 6.9 \times 10^{-4}\, \mathrm{eV}^{2}$ and $\vartheta_{V} = 16.5^{\circ}$. The density and mixing parameters are chosen so as to put the $q = 1\, \mathrm{MeV}$ neutrinos on an MSW resonance. From these parameters we find the wavelength of the oscillations of the neutrinos to be $6.6\;{\rm km}$ and for the antineutrinos it is $2\;{\rm km}$.
All species of neutrinos and antineutrinos are assumed to be emitted half-isotropically (i.e., $\beta=0$) from a neutrinosphere at $R_{\nu}=10 \; \mathrm{km}$.

\paragraph*{Multi-angle code —}
To perform the steady-state multi-angle calculation, the emission angle relative to the radial direction at the neutrinosphere, $\theta_R$, is discretized into $N_A = 9001$ angle bins uniformly spaced in $u_R = \sin^2 \theta_R$. This distribution in angles ensures a better sampling of the path lengths from the neutrinosphere to a given radial point $r$ compared to angle bins uniformly sampled in $\theta_R$. 
The evolution matrices $S$ and $\bar{S}$ are parameterized to ensure unitarity and the evolution equations for the parameters are solved using an explicit adaptive step size Runge-Kutta algorithm employing the Cash-Karp set of coefficients. The radial step size is limited to allow a maximum absolute numerical error of the integration variables of $10^{-10}$. At any given radial coordinate $r$ we construct the moments according to the definitions given in Eqs.~\eqref{eq:defE} through \eqref{eq:defP}. We can then compute the `true' closure parameters from these constructed moments.

\paragraph*{Moment code —}
The steady-state moment evolution code we use is a slightly modified version of the code described in~\cite{Myers:2021hnp}. The small modification we have made is to use the same adaptive step-size Runge Kutta algorithm as the multi-angle code described above to solve the ODEs for the moments rather than a partial differential equation solver. 

To solve the moment evolution equations we need to provide a closure. The exact closure is known already from the results of the steady-state multi-angle code but in what follows, we shall consider cases where we make only partial use of the closure parameters.

\subsubsection{Results}

\begin{figure*}[!ht]
    \includegraphics[width=0.85\linewidth]{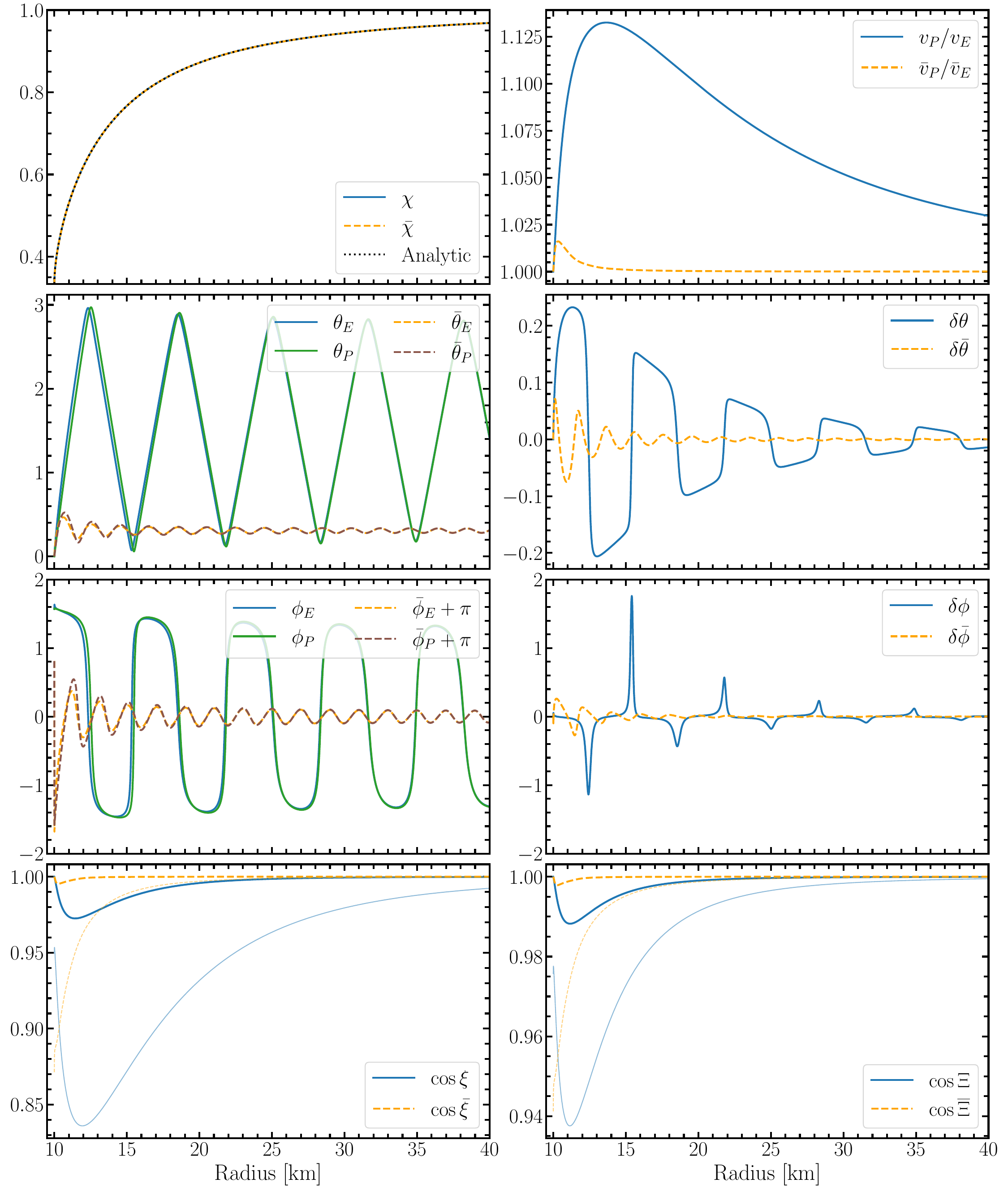} 
\caption{Parameters of the quantum closure as a function of the radial coordinate for the constant matter test problem described in Sec.~\ref{sec:MSWproblem}: $\chi$ and ${\bar{\chi}}$ (top left panel), ratios $v_P/v_E$, $\bar{v}_P/\bar{v}_E$ (top right panel), angles $\theta_E$, $\theta_P$, ${\bar{\theta}}_E$ and ${\bar{\theta}}_P$ (second row, left panel), their differences $\delta \theta = \theta_E - \theta_P$ and $\delta \bar{\theta} = \bar{\theta}_E - \bar{\theta}_P$ (second row, right panel), phases $\phi_E$, $\phi_P$, ${\bar{\phi}}_E$ and ${\bar{\phi}}_P$ (third row, left panel), their differences $\delta \phi = \phi_E - \phi_P$ and $\delta \bar{\phi} = \bar{\phi}_E - \bar{\phi}_P$ (third row, right panel), and the alignment angles $\cos \xi$, $\cos \bar{\xi}$ computed from Eq.~\eqref{eq:cosxi} (bottom left panel) and $\cos \Xi$, $\cos \bar{\Xi}$ computed from Eq.~\eqref{eq:Xi} (bottom right panel). In the top left panel we also plot the analytic expression for $\chi$ and ${\bar{\chi}}$ given in Eq.~\eqref{eq:analytic_chi}. We added $\pi$ to the antineutrino phases $(\bar{\phi}_E, \, \bar{\phi}_P)$ on the third row, left panel, so that they vary in the same range as $(\phi_E,\,\phi_P)$. In the bottom panels, the light curves show the lower limits on the alignment angles given by the right-hand side of Eqs.~\eqref{eq:cosXilimit} and \eqref{eq:cosxilimit0}.} 
\label{fig:MSWclosure}
\end{figure*}

\paragraph*{Multi-angle calculation ---} We first show on Fig.~\ref{fig:MSWclosure} the closure parameters obtained from the multi-angle solution of this MSW test problem. Since the two neutrino flavors have the same neutrinosphere, we are able to compute the $\chi$ parameter analytically in this case~\cite{Myers:2021hnp} and find:
\begin{equation}
\label{eq:analytic_chi}
    \chi(r) = \frac13 \left[2 - \left(\frac{R_\nu}{r}\right)^2 + \sqrt{1 - \left(\frac{R_\nu}{r}\right)^2}\right] \, .
\end{equation}
This expression is shown as the black dotted line on Fig.~\ref{fig:MSWclosure}, and agrees very closely with the multi-angle results. As we get away from the neutrinosphere at $R_\nu = 10 \, \mathrm{km}$, $\chi$ increases from $1/3$ — consistent with the half-isotropic emission at $r= R_\nu$ — to $1$ — as the trajectories of the neutrinos become more closely aligned with the radial direction.

The evolution of the other closure parameters can be understood as a consequence of the decreasing coherence of neutrinos which have traveled along different trajectories from the neutrinosphere to a given point above it, and the different weighting — the $\hat{q}$ terms in Eqs.~\eqref{eq:defE} through \eqref{eq:defP} for the definition of the moments — of those trajectories when computing the moments. For a single neutrino or antineutrino the eigenvalues of the density matrix do not change along the trajectory because the eigenstates of a constant Hamiltonian (as in our case) have constant amplitude, but the eigenvalues of a moment converge due to decoherence. The $P_{rr}$ pressure moment weights the neutrinos traveling along the radial direction more heavily than neutrinos traveling along other directions, whereas the energy density weights them all equally. Thus initially the $v_P$ parameter, which measures the splitting of the eigenvalues of $P_{rr}$, decreases more slowly than $v_E$ and so the ratio $v_P/v_E$ initially increases. But $v_E$ and $v_P$ do not decrease with the radius $r$ to zero in this particular problem: there is always some finite amount of coherence between neutrinos because the path-length difference of any two neutrinos is bounded from above by the radius of the neutrinosphere which is $R_{\nu} = 10\;{\rm km}$, and the oscillation wavelength of a neutrino — $6.6\;{\rm km}$ with our chosen parameters — is similar to this scale. Thus eventually both $v_E$ and $v_P$ will approach a finite asymptote and we should expect $v_E$ to approach the asymptote at a smaller radius $r$ than $v_P$. Therefore once $v_E$ is sufficiently close to its asymptotic value, the ratio $v_P/v_E$ will decrease due to the continued evolution of $v_P$. The same explanation also works for the antineutrinos but now the oscillation wavelength is only $2\;{\rm km}$, so the turnover of the ratio $\bar{v}_P / \bar{v}_E$ happens at a smaller radius $r$.

The closure parameters $\theta_E$, $\theta_P$, $\phi_E$ and $\phi_P$ measure the orientation of the `spatial' of the moment 4-vectors in flavor space part — see Appendix~\ref{sec:app1}. Note that the wavelength of the oscillations of $\phi_{E,P}$ (resp. $\bar{\phi}_{E,P}$) is consistent with the MSW oscillation length of $6.6 \, \mathrm{km}$ (resp. $2 \, \mathrm{km}$). For a single neutrino the equivalent ‘spatial’ part of the density matrix precesses around the spatial part of the Hamiltonian~\cite{Blennow:2013rca}. The closer the neutrino is to the MSW resonance, the more the axis of precession lies in the $(\widetilde{x},\widetilde{y})$ plane. The moments behave similarly, except that the angle of the precession cone is reduced because of decoherence, since we sum contributions from the spatial parts of individual neutrino density matrices which have different phases.

Finally, the two angles $\xi$ and $\Xi$ measure the alignment of the moments. The evolution of these angles with radius $r$ is similar to the evolution of  $v_P/v_E$ and $\bar{v}_P/\bar{v}_E$: initially the angle between the moments grows but then reaches a maximum — $\xi_{max} \sim 14^{\circ}$, $\Xi_{max} \sim 8^{\circ}$ for neutrinos and much smaller for antineutrinos — and then turnover and approach zero at large radii. Again, this behavior is to be expected due to the different rates at which the moments approach their asymptotic solutions. The light curves on the bottom panel of Fig.~\ref{fig:MSWclosure} show the minimum values allowed for $\cos \Xi$ and $\cos \xi$ according to Eqs.~\eqref{eq:cosXilimit} and \eqref{eq:cosxilimit0}. The alignment angles are always significantly smaller than the allowed limit, except at large radii where decoherence ‘locks’ the energy and pressure moments, such that the limits asymptote to 1.

\begin{figure}[!ht]
    \includegraphics[width=\linewidth]{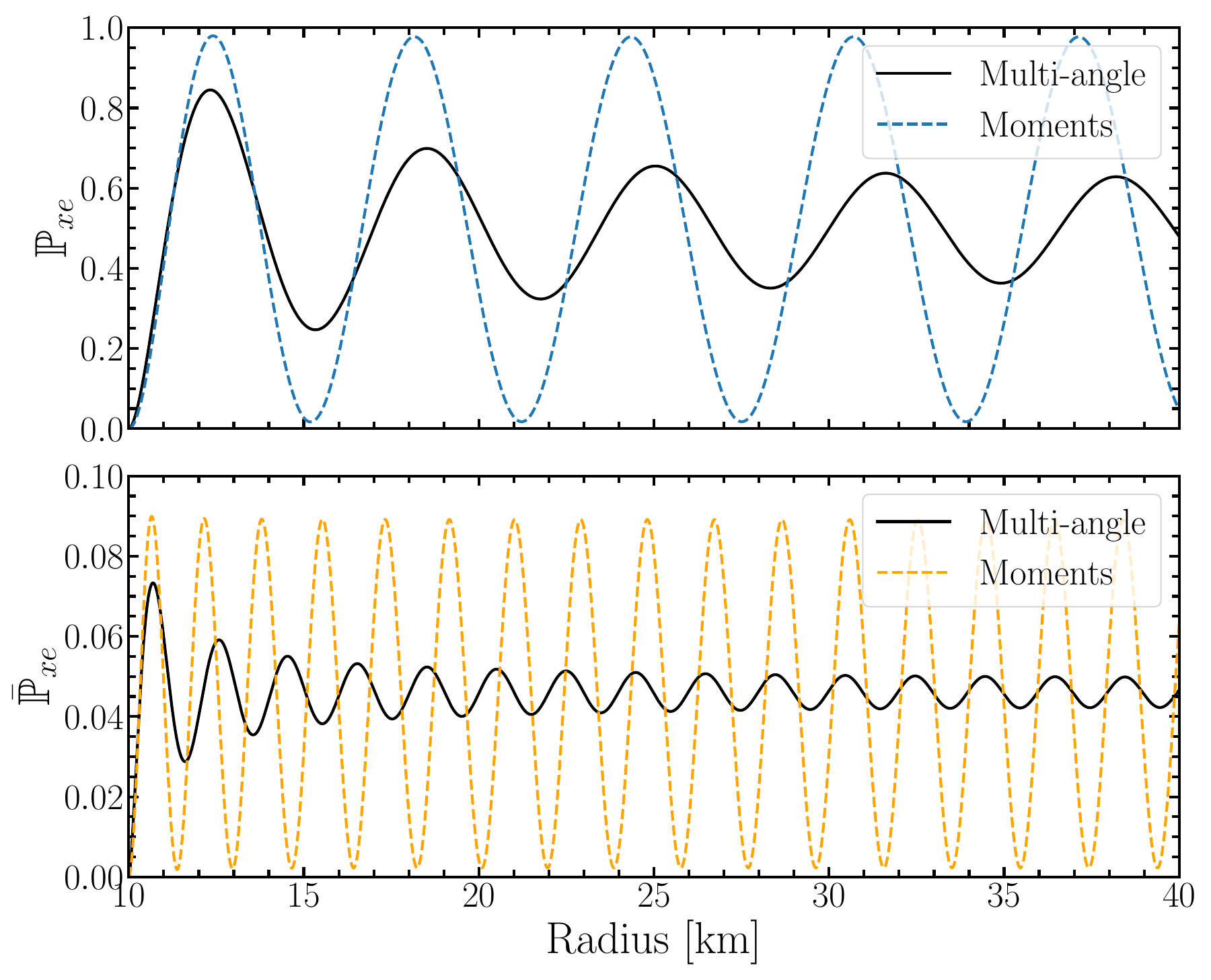}
\caption{Net transition probability $\mathbb{P}_{xe}$ for neutrinos (top panel), and $\bar{\mathbb{P}}_{xe}$ for antineutrinos (bottom panel) as a function of the radial coordinate, computed from the moment and multi-angle codes. $\chi$ and ${\bar{\chi}}$ are computed from the multi-angle calculation then used as input to the moment code. All the other closure parameters are assigned as $v_E = v_P$, ${\bar{v}}_E = {\bar{v}}_P$, $\theta_E = \theta_P$, ${\bar{\theta}}_E = {\bar{\theta}}_P$, $\phi_E = \phi_P$, and ${\bar{\phi}}_E = {\bar{\phi}}_P$.}
\label{fig:MSWQC1}
\end{figure}

\begin{figure}[!ht]
    \includegraphics[width=\linewidth]{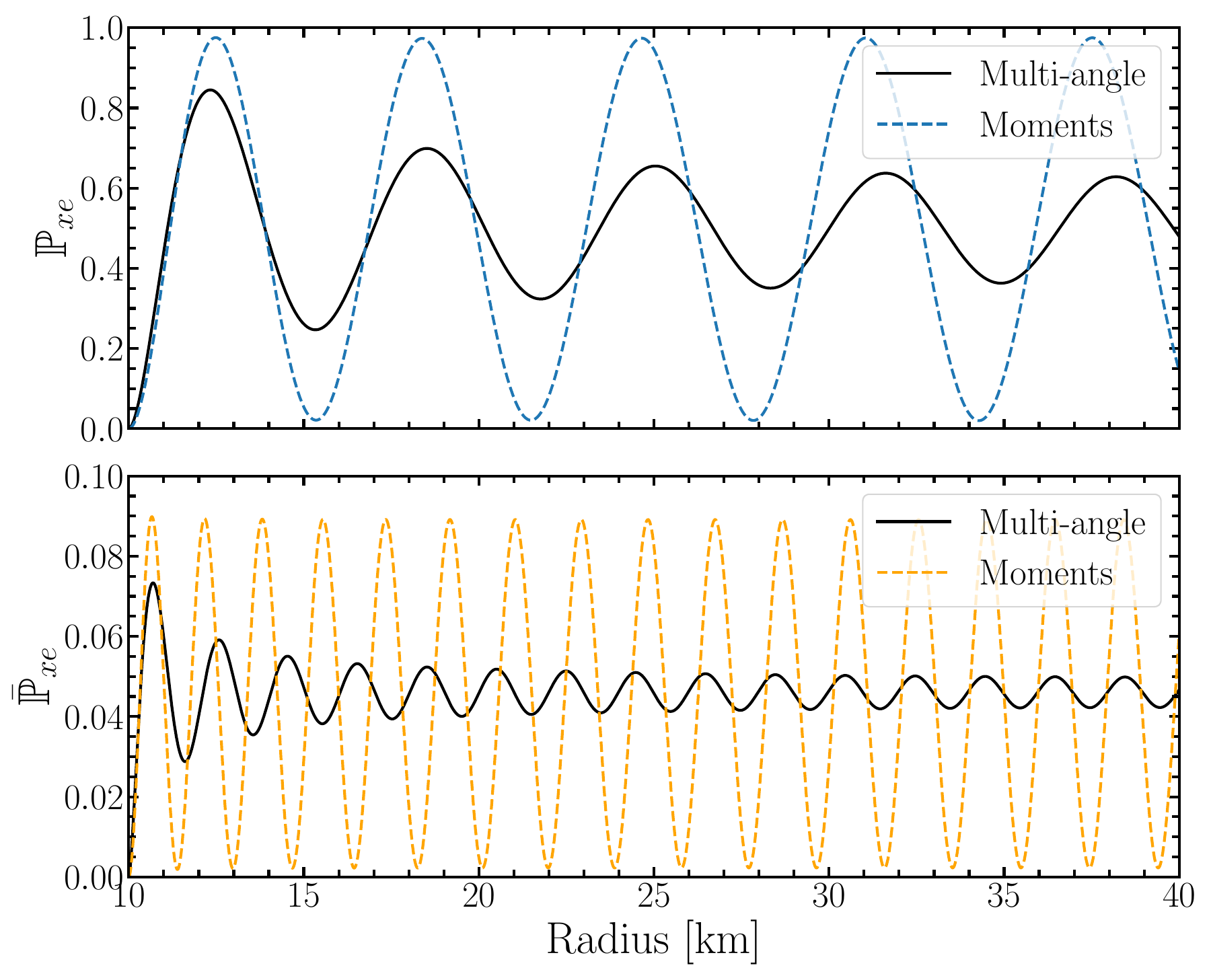}
\caption{The same as Fig.~\ref{fig:MSWQC1} but now we also compute $\delta v = v_E - v_P$ and $\delta{\bar{v}} = {\bar{v}}_E - {\bar{v}}_P$ from the multi-angle results (see Fig.~\ref{fig:MSWclosure}) and in the moment code assign $v_E = v_P + \delta v$ and ${\bar{v}}_E = {\bar{v}}_P + \delta{\bar{v}}$.}
\label{fig:MSWQC2}
\end{figure}

\begin{figure}[!ht]
    \includegraphics[width=\linewidth]{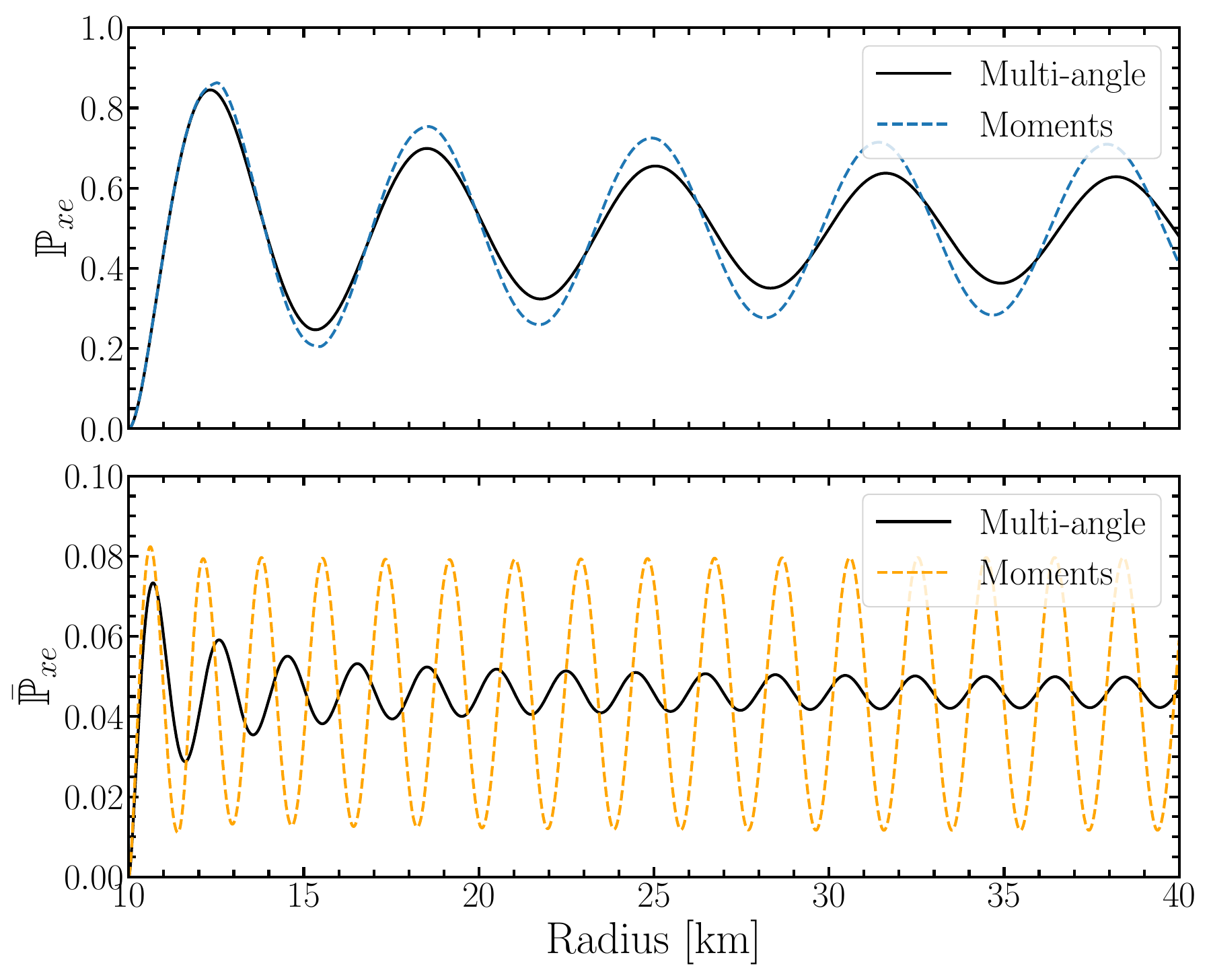}
\caption{The same as Fig.~\ref{fig:MSWQC2} but now we also compute $\delta\theta = \theta_E - \theta_P$ and $\delta{\bar{\theta}} = {\bar{\theta}}_E - {\bar{\theta}}_P$ from the multi-angle results and in the moment code assign $\theta_E = \theta_P + \delta\theta$ and ${\bar{\theta}}_E = {\bar{\theta}}_P + \delta{\bar{\theta}}$ in the closure.}
\label{fig:MSWQC3}
\end{figure}

\begin{figure}[!ht]
    \includegraphics[width=\linewidth]{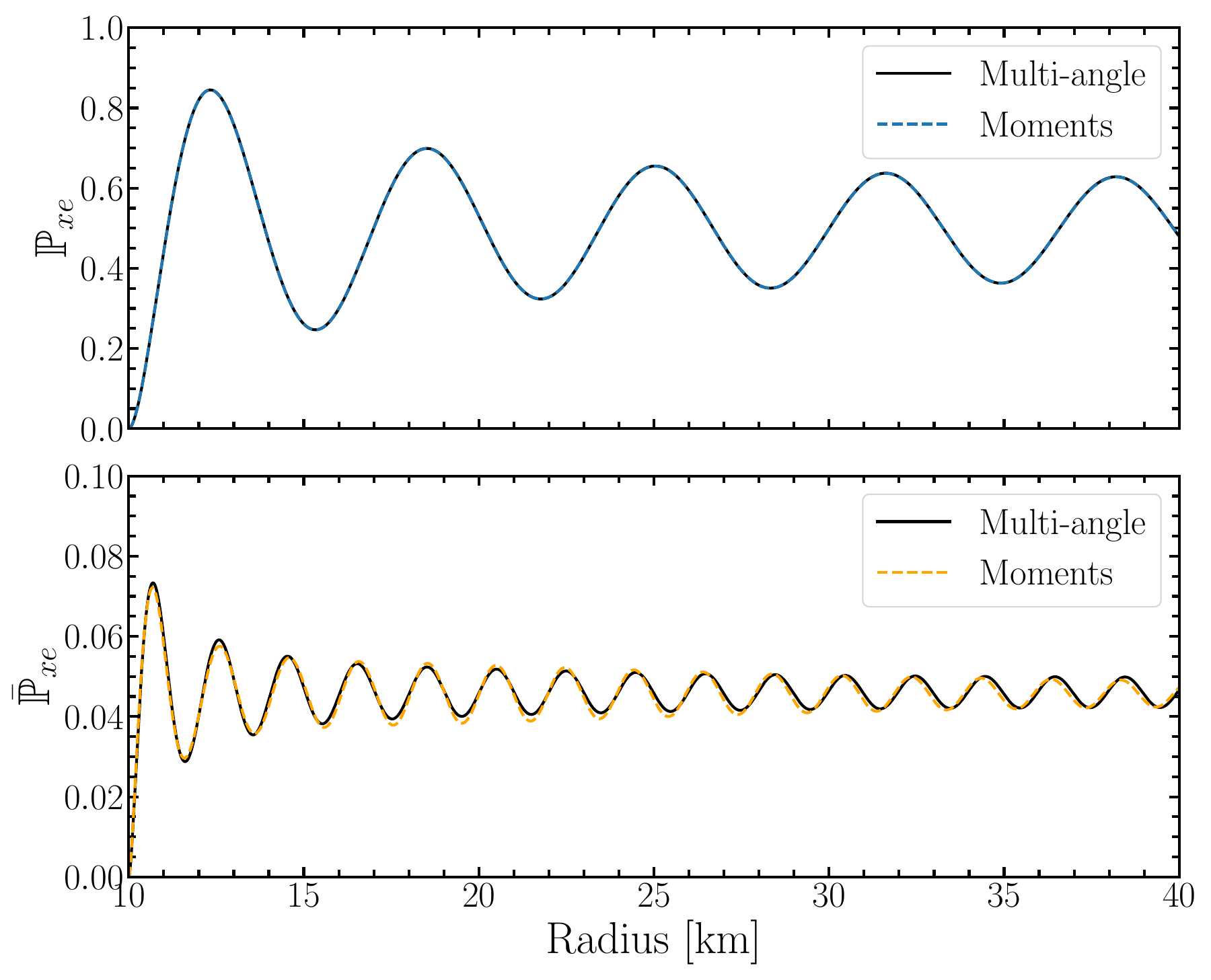}
\caption{The same as Fig.~\ref{fig:MSWQC3} but we now extract $\delta\phi = \phi_E - \phi_P$ and $\delta{\bar{\phi}} = {\bar{\phi}}_E - {\bar{\phi}}_P$ from the multi-angle results and assign $\phi_E = \phi_P + \delta\phi$ and ${\bar{\phi}}_E = {\bar{\phi}}_P + \delta{\bar{\phi}}$ in the closure of the moment code.}
\label{fig:MSWQC4}
\end{figure}

\paragraph*{Moment calculation with quantum closure —} In Figures \ref{fig:MSWQC1} through \ref{fig:MSWQC4} we show the results of the moment code for four different closures. The plotted quantity is the effective transition probability defined from the flux moment to be 
\begin{align}
\mathbb{P}_{\nu_x \rightarrow \nu_e}(r) \equiv \mathbb{P}_{xe} &= \frac{r^2\,F_{r}^{(ee)}(r) - R_{\nu}^2\,F_{r}^{(xx)}(R_{\nu})}{R_{\nu}^2\,[F_{r}^{(ee)}(R_{\nu}) - F_{r}^{(xx)}(R_{\nu})]} \, , \label{eq:Ptrans}\\
\mathbb{P}_{\bar{\nu}_x \rightarrow \bar{\nu}_e}(r) \equiv \bar{\mathbb{P}}_{xe} &=  \frac{r^2\,\bar{F}_{r}^{(ee)}(r) - R_{\nu}^2\,\bar{F}_{r}^{(xx)}(R_{\nu})}{R_{\nu}^2\,[ \bar{F}_{r}^{(ee)}(R_{\nu}) - \bar{F}_{r}^{(xx)}(R_{\nu})] } \label{eq:Ptransbar} \, .
\end{align}

In Fig.~\ref{fig:MSWQC1} we use a closure where we only use the closure parameters $\chi(r)$ and ${\bar{\chi}}(r)$ computed from the multi-angle results — which are, once again, analytically predictable with Eq.~\eqref{eq:analytic_chi} — while the remaining closure parameters are chosen to be $v_E = v_P$, ${\bar{v}}_E = {\bar{v}}_P$, $\theta_E = \theta_P$, ${\bar{\theta}}_E = {\bar{\theta}}_P$, $\phi_E = \phi_P$, and ${\bar{\phi}}_E = {\bar{\phi}}_P$. This is the same case as \cite{Myers:2021hnp}. From the figure we see that while the wavelength of the flavor oscillations of both the neutrinos and antineutrinos is approximately correct, the amplitude is clearly wrong when using this closure. This moment calculation is not able to fully account for the decoherence of the neutrinos traveling along different trajectories from the neutrinosphere which results in a decrease of the amplitude of $(\mathbb{P}_{xe}, \, \bar{\mathbb{P}}_{xe})$ with increasing radius.

In Fig.~\ref{fig:MSWQC2} we again use closure parameters $\chi(r)$ and ${\bar{\chi}}(r)$ computed from the multi-angle results and also compute $\delta v = v_E - v_P$ and $\delta{\bar{v}}={\bar{v}}_E - {\bar{v}}_P$. In the moment code we compute $v_P$ and ${\bar{v}}_P$ from $P_{rr}$ and ${\bar{P}}_{rr}$ using Eq.~\eqref{eq:components_P} and the parameters $v_E$ and ${\bar{v}}_E$ are obtained as $v_E = v_P + \delta v$ and ${\bar{v}}_E = {\bar{v}}_P + \delta {\bar{v}}$. The remaining closure parameters remain as $\theta_E = \theta_P$, ${\bar{\theta}}_E = {\bar{\theta}}_P$, $\phi_E = \phi_P$, and ${\bar{\phi}}_E = {\bar{\phi}}_P$. Since $\delta v$ and $\delta {\bar{v}}$ are non-zero, this quantum closure allows different flavors to be treated differently. Compared to Fig.~\ref{fig:MSWQC1}, we observe a moderate improvement of the agreement of the wavelength of the flavor oscillations.

For the next iteration of the quantum closures we start with the closure used to make Fig.~\ref{fig:MSWQC2} and replace the assignment $\theta_E = \theta_P$, ${\bar{\theta}}_E = {\bar{\theta}}_P$ by computing the quantities $\delta \theta = \theta_E - \theta_P $ and $\delta{\bar{\theta}} = {\bar{\theta}}_E - {\bar{\theta}}_P$ from the multi-angle results. We find $\theta_E$ and ${\bar{\theta}}_E$ by first finding $\theta_P$ and ${\bar{\theta}}_P$ from $P_{rr}$ and ${\bar{P}}_{rr}$ and then using $\theta_E = \theta_P + \delta \theta$ and ${\bar{\theta}}_E = {\bar{\theta}}_P + \delta {\bar{\theta}}$. Fig.~\ref{fig:MSWQC3} shows the results and we observe that the agreement of the amplitude of the flavor oscillations of the neutrinos is improved compared to the results seen in Fig.~\ref{fig:MSWQC2}, but the amplitude of the flavor oscillations of the antineutrinos barely changes from that in Fig.~\ref{fig:MSWQC2}.

Finally, in Fig.~\ref{fig:MSWQC4} we replace the assignment in the closure of $\phi_E = \phi_P$, and ${\bar{\phi}}_E = {\bar{\phi}}_P$ with $\phi_E = \phi_P + \delta\phi$, and ${\bar{\phi}}_E = {\bar{\phi}}_P + \delta{\bar{\phi}}$ where $\delta\phi = \phi_E - \phi_P$ is taken from the multi-angle results and similarly for $\delta{\bar{\phi}} = {\bar{\phi}}_E - {\bar{\phi}}_P$. The figure shows that the results from the moment code and the multi-angle code now match very well. The small differences are attributable to numerical error coming from the ODE integrator and/or the algorithms used to interpolate $\delta v$, $\delta{\bar{v}}$, $\delta\theta$, $\delta{\bar{\theta}}$, $\delta\phi$ and $\delta{\bar{\phi}}$.

\subsection{Homogeneous Fast Flavor Instability}

We now consider a system where self-interactions drive the dynamics, specifically, a system experiencing a “fast” flavor instability~\cite{Richers_review}. 

\subsubsection{Numerical setup}

We consider a homogeneous, mono-energetic gas of neutrinos and antineutrinos in cartesian coordinates with axial symmetry around the $z$ axis. The equations of motion are the same as \eqref{eq:QKE_multi} and \eqref{eq:QKE_moments} with $r \to z$, and without the spatial derivative and geometric terms coming from spherical symmetry on the left-hand side. Namely, it reads for neutrinos:
\begin{align}
    \frac{\partial \varrho}{\partial t} &= - \imath \left[H_V + H_M + H_E - \mu H_F, \varrho\right] \, , \label{eq:QKE_multi_hom} \\
    \frac{d E}{d t}  &= - \imath \left[H_V + H_M + H_E, E\right] + \imath \left[H_F, F_z \right] \, , \\
    \frac{d F_z}{d t}  &= - \imath \left[H_V + H_M + H_E, F_z \right] + \imath \left[H_F, P_{zz} \right] \, .
\end{align}
With our choice of initial conditions, this system presents a crossing between the electron neutrino and antineutrino angular distributions, hence it should present a fast flavor instability (FFI)~\cite{Morinaga_Crossing_2022,dasgupta_CollectiveNeutrinoFlavor_2022}.  In this particular case, the system is homogeneously unstable, so we can demonstrate evolution of a flavor instability without considering inhomogeneous perturbations.

The first angular moments of the (anti)neutrino initial distributions we consider are given in Table~\ref{tab:initial_FFI}. They correspond to the conditions found in an actual neutron star merger environment. Specifically, the flavor configuration is obtained from the 5 ms post-merger snapshot of the M1 simulation in~\cite{Foucart:2016rxm}, namely, from the grid point of coordinates $(X=64.6 \, \mathrm{km}, \, Y=0 \, \mathrm{km}, \, Z = 6.8 \, \mathrm{km})$ — see, e.g.,~\cite{Froustey:2023skf,Richers:2024zit} for a discussion of the FFI in this NSM snapshot. The local basis $(x,y,z)$ is chosen such that the net ELN flux $\vec{F}^{(ee)} - \vec{\bar{F}}^{(ee)}$ is along the $z$ axis. With this choice, the $x,y$ components of $\vec{F}$ for each species are less than $10 \, \%$ of the $z$ component, and we fully neglect these transverse parts to obtain an axisymmetric system. We can compute the average energy of each species from the results of~\cite{Foucart:2016rxm}, which leads to $\langle q \rangle_{\nu_e} \simeq \langle q \rangle_{\bar{\nu}_{e}} \simeq 16 \, \mathrm{MeV}$ and $\langle q \rangle_{\nu_x} \simeq 28 \, \mathrm{MeV}$ for the point we consider. We neglect this difference here, and consider a single energy bin with the average energy $q_0 = 22 \, \mathrm{MeV}$. The energy and number density moments of each species are then proportional, $E = q_0 N$. Finally, note that the M1 scheme in \cite{Foucart:2016rxm} considers three species ($\nu_e$, $\bar{\nu}_e$ and $\nu_X$ combining $\nu_\mu$, $\nu_\tau$, $\bar{\nu}_\mu$ and $\bar{\nu}_\tau$), such that in our two-flavor framework we have equal initial distributions of $\nu_x$ and $\bar{\nu}_x$.

\renewcommand{\arraystretch}{1.2}

\begin{table}[!ht]
    \centering
    \begin{tabular}{c|ccc}
    Species & $N \, (10^{33} \, \mathrm{cm}^{-3})$ & $\ \, F_z/E \ \, $ & $\ \, P_{zz}/E \ \, $ \\ \hline
    $\nu_e$ & $1.525$ & $0.114$ & $0.338$ \\
    $\nu_x$ & $0.134$  & $0.168$ & $0.344$ \\
    $\bar{\nu}_e$ & $1.301$ & $0.221$ & $0.352$ \\
    $\bar{\nu}_x$ & $0.134$ & $0.168$ & $0.344$
    \end{tabular}
    \caption{Initial moments for the homogeneous FFI test calculation, corresponding to the point located at $(X=64.6 \, \mathrm{km}, \, Y=0 \, \mathrm{km}, \, Z = 6.8 \, \mathrm{km})$ in the 5 ms post-merger snapshot from~\cite{Foucart:2016rxm}. The pressure moments are calculated from the Minerbo closure [see Eq.~\eqref{eq:chi_ME}].}
    \label{tab:initial_FFI}
\end{table}

At this grid point in the simulation, the matter density is $\rho = 7.23\times10^{10} \, \mathrm{g.cm^{-3}}$, and the electron fraction is $Y_e = 0.224$, such that the electron number density is $n_e = 9.70 \times 10^{33} \, \mathrm{cm^{-3}}$. In the vacuum term, we consider a mass-squared difference $\Delta m^2 = 2.5 \times 10^{-3} \, \mathrm{eV}^2$ and a mixing angle $\vartheta_V = 0.587 \, \mathrm{rad}$ (corresponding to $\vartheta_{12}$ in the 3-flavor framework~\cite{ParticleDataGroup:2024cfk}).\footnote{Given the very large matter term compared to the vacuum contribution, this configuration is equivalent to an effective mixing angle $\vartheta_M \simeq 2 \times 10^{-8} \, \mathrm{rad}$ and an effective mass-squared difference $\Delta m_M^2 \simeq 5.4 \times 10^4 \, \mathrm{eV}^2$.}

\paragraph*{Multi-angle code —}
The homogeneous QKEs are written as a set of $2 \times N_A$ ODEs, with $N_A$ the number of angular bins. For this calculation, we use $N_A = 40$. The differential equations are solved with the \texttt{solve\_ivp} function from the \texttt{SciPy} package in Python~\cite{Scipy}, with tolerance parameters $\mathtt{rtol} = 10^{-11}$ and $\mathtt{atol} = 10^{-12}$ to ensure sufficient precision.

The initial angular distributions corresponding to the moments given in Table~\ref{tab:initial_FFI} are determined by the classical maximum entropy closure (or “Minerbo” closure)~\cite{Minerbo_1978,Cernohorsky_closure_1994} used in the simulation~\cite{Foucart:2016rxm} from which we extract this initial flavor configuration. The maximum entropy distribution is:
\begin{equation}
\label{eq:distrib_ME}
    f(\mu \, ;N,F_z) = \frac{N}{4 \pi} \frac{Z}{\sinh(Z)} e^{Z \mu} \, ,
\end{equation}
where $Z$ is solution of the equation
\begin{equation}
\label{eq:Z_ME}
    \frac{F_z}{E} = \coth(Z) - \frac{1}{Z} \, .
\end{equation}
The pressure moment is then given by $P_{zz}=E-2F_z/E = \chi(F_z/E) E$. The Eddington factor is given at 2~\% accuracy by the polynomial form~\cite{Cernohorsky_closure_1994,Smit_closure_2000}:
\begin{equation}
\label{eq:chi_ME}
    \chi(\hat{f}) = \frac{1}{3} + \frac{2 \hat{f}^2}{15}\left(3-\lvert\hat{f}\rvert+3 \hat{f}^2\right) \, .
\end{equation}
The initial density matrices are diagonal in flavor space: at each angular bin $\mu_n$, we set $\varrho^{(aa)}(\mu_n,t=0) = f(\mu_n \, ;N^{(aa)},F_{z}^{(aa)})$, with $N^{(aa)}$ and $F_{z}^{(aa)}$ the number density and flux density of $\nu_a$ given in Table~\ref{tab:initial_FFI}. The flavor off-diagonal components become non-zero because of the vacuum term, which seeds the fast flavor instability.

\paragraph*{Moment code ---} We solve the homogeneous moment QKEs with the same solver as in the multi-angle case (\texttt{solve\_ivp} function from the \texttt{SciPy} library~\cite{Scipy}). The initial angular moments $(N,F^z)$ are set by the values in Table~\ref{tab:initial_FFI}, and we adopt different closures based on parameters extracted from the multi-angle results.

\subsubsection{Results}

\begin{figure*}
    \centering
    \includegraphics[width=0.95\linewidth]{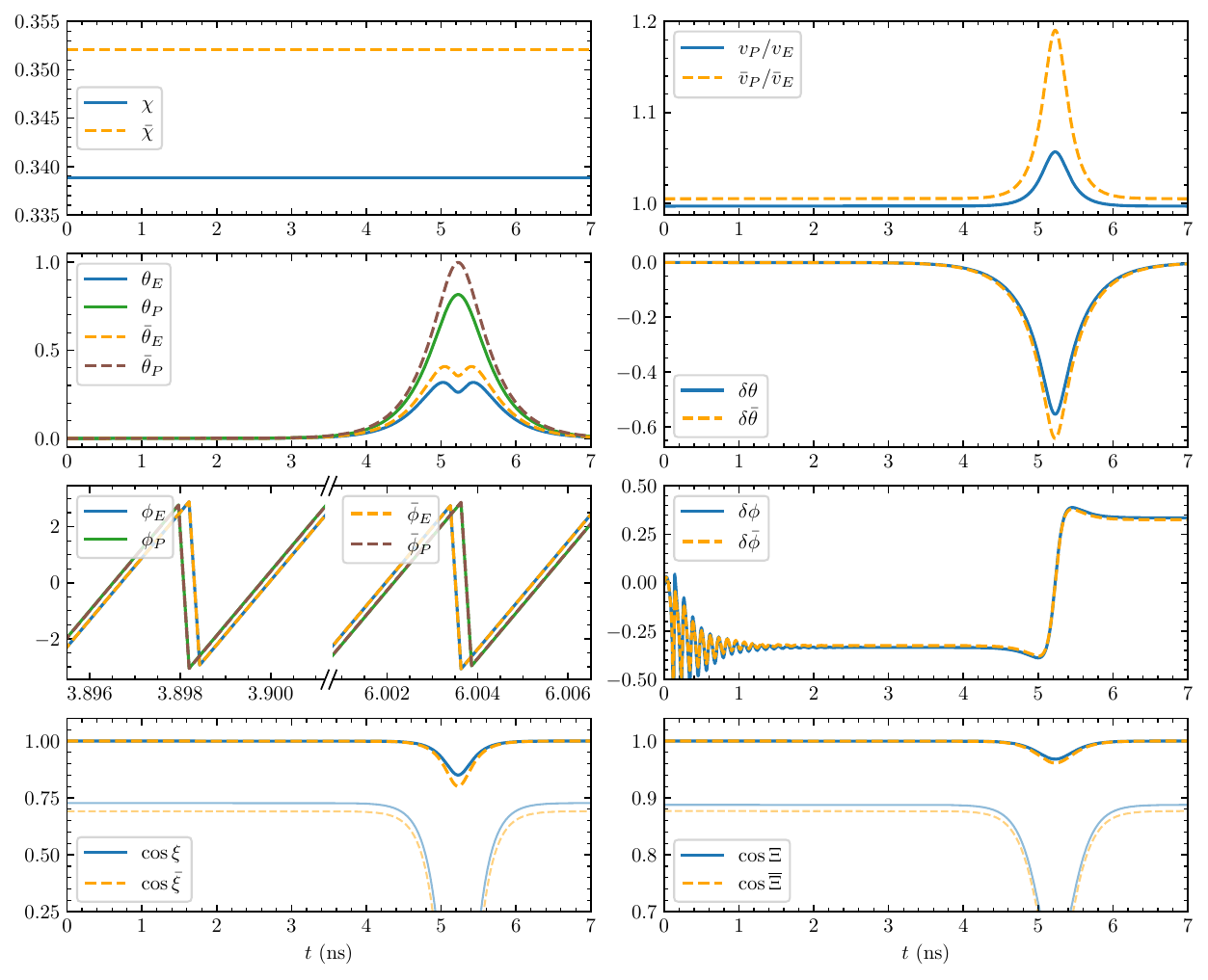}
    \caption{Closure parameters and alignment angles between the $E$ and $P_{rr}$ moments from the multi-angle solution of the homogeneous FFI test case. (Anti)neutrino quantities are in solid (dashed) lines. The left panel on the third row show the phases, which change on a timescale set by the matter Hamiltonian, before and after the peak of the instability. In the bottom panels, the light curves show the lower limits on the alignment angles given by the right-hand side of Eqs.~\eqref{eq:cosXilimit} and \eqref{eq:cosxilimit0}.} 
    \label{fig:FFIclosure}
\end{figure*}

\paragraph*{Multi-angle calculation ---} The closure parameters and alignment angles are shown on Fig.~\ref{fig:FFIclosure}, similarly to Fig.~\ref{fig:MSWclosure}. 
Given the structure of the QKEs~\eqref{eq:QKE_multi_hom} for an homogeneous system, the traces of $\varrho$ and therefore of all angular moments are conserved. This means that $E^{\, \widetilde{t}}$ and $P_{rr}^{\, \widetilde{t}}$ are conserved quantities, such that $\chi$ is also constant according to Eq.~\eqref{eq:Pt_Et}. This is consistent with the top left panel of Fig.~\ref{fig:FFIclosure}. The higher value of $\bar{\chi}$ is due to the higher initial flux factor for $\bar{\nu}_e$ compared to $\nu_e$. The FFI manifests itself by the exponential growth of the angles $\theta_{E,P}, \, \bar{\theta}_{E,P}$, until saturation is reached for $t_\mathrm{sat} \sim 5.3 \, \mathrm{ns}$. On the third row, we show the phases $\phi_{E,P}, \, \bar{\phi}_{E,P}$ and their differences. On the left panel, we zoomed in on two time domains corresponding to pre- and post-saturation of the instability, where the phase difference goes from $\delta \phi \simeq -0.334 \, \mathrm{rad}$ to $\delta \phi \simeq +0.334 \, \mathrm{rad}$ (and $\delta \bar{\phi} \simeq -0.324 \, \mathrm{rad}$ to $\delta \bar{\phi} \simeq +0.324 \, \mathrm{rad}$), see right panel. Note the very short timescale of variation of the phases, which is set by the matter Hamiltonian. For $t < 1.8 \, \mathrm{ns}$, we see large oscillations in the phase differences: this corresponds to a transient regime as the fastest growing mode emerges. Once the most unstable mode dominates, the values of the parameters match the features of the mode with the largest growth rate that can be found in linear stability analysis (see, e.g.,~\cite{Froustey:2023skf}).

\begin{figure}[ht]
    \centering
    \includegraphics[width=\columnwidth]{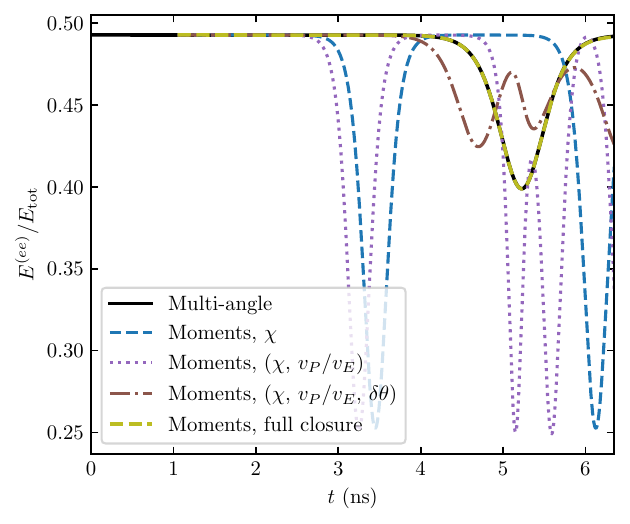}
    \caption{Evolution of the electron neutrino energy density, for different implementations of the quantum closure. In dashed blue, the parameters $\chi$ and $\bar{\chi}$ from the multi-angle results are used in the moment calculation, the other parameters being set to $v_P=v_E$, $\delta \theta = 0$ and $\delta \phi = 0$ (and likewise for antineutrinos). These other parameters are progressively included in the other calculations: $v_P/v_E$ in dashed purple, $\delta \theta$ in dash-dotted brown and finally $\delta \phi$ (i.e., the full closure) in dashed green.}
    \label{fig:FFI_ee}
\end{figure}

\begin{figure}[ht]
    \centering
    \includegraphics[width=\columnwidth]{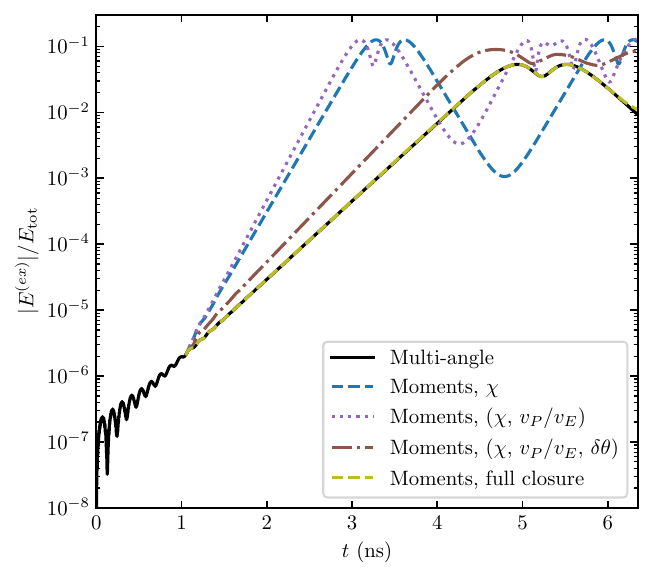}
    \caption{Evolution of the flavor off-diagonal component of the energy density matrix, for different implementations of the quantum closure (see caption of Fig.~\ref{fig:FFI_ee}).}
    \label{fig:FFI_offdiag}
\end{figure}

\paragraph*{Moment calculation with quantum closure ---} Similarly to the steady-state calculation in Sec.~\ref{sec:MSWproblem}, we compare the results of solving the moment equations by imposing progressively more and more elements of the closure — when all parameters are imposed, if the parameterization is complete, the results of the moment calculation should agree with the multi-angle ones. We show the evolution of the electron neutrino energy density (normalized by the conserved total energy density $E_\mathrm{tot} = E^{(ee)} + E^{(xx)} + \bar{E}^{(ee)} + \bar{E}^{(xx)}$) on Fig.~\ref{fig:FFI_ee}, and the magnitude of the flavor off-diagonal component $\lvert E^{(ex)} \rvert$ on Fig.~\ref{fig:FFI_offdiag}. The `true' results of the multi-angle calculation are shown with a solid black line. When only the $\chi, \, \bar{\chi}$ values are used (dashed blue lines), the growth rate is over-estimated such that the conversion $\nu_e \to \nu_x$ occurs earlier ($t \sim 3 \, \mathrm{ns}$). Interpolating the actual value of $v_P/v_E$ does not improve the results, while accounting for the difference in growth of the $\widetilde{x}, \, \widetilde{y}$ components of the moments (through $\delta \theta, \, \delta \bar{\theta}$) gives more accurate results (see dash-dotted brown line). Finally, and as expected, using the full closure allows to recover the multi-angle results.

We do not start the moment calculation at $t=0$ but at a later stage of the development of the instability, because of the extremely fast variations of $\delta \phi$ in initial time steps (see Fig.~\ref{fig:FFIclosure}, third row, right panel). Tracking accurately these first time steps would require an enormous time resolution — if it is not sufficient, we accumulate small errors which blow up in the decoherence phase, driving premature numerical divergence of the moment calculation from the multi-angle calculation. In addition, the very short timescales associated to the matter part of the Hamiltonian make the calculation much more challenging.

The implementation shown in dashed blue is equivalent to setting the pressure proportional to the energy density for all flavor matrix entries, i.e., $P_{zz}^{(a b)} = \chi E^{(ab)}$, where $\chi$ is determined by the flavor trace of $P_{zz}$ and $E$. This approach is similar to the one used in the moment calculations of Ref.~\cite{Grohs:2022fyq,Grohs:2023pgq}, where faster growth rates were consistently found compared to multi-angle calculations, a trend also verified here. 

It it worth noting that the implementations where only some parts of the quantum closure are used are not rigorously self-consistent. Indeed, the parameters ($v_P/v_E$, $\delta \theta$, $\delta \phi$...) are interpolated as a function of time, and the interpolated values are used as input to calculate the pressure tensor in Eqs.~\eqref{eq:flux}--\eqref{eq:fluxbar}. However, as can be clearly seen on Fig.~\ref{fig:FFI_offdiag}, an incomplete closure leads here to larger growth rates. As a consequence, if we take for instance the dash-dotted brown curves, the maximum value of $\theta_E$ (which corresponds to the peak of $\lvert E^{(ex)} \rvert$) is attained earlier than in the actual evolution. We see on Fig.~\ref{fig:FFIclosure} that $\delta \theta(t)$ decreases until $t_\mathrm{sat}$, then increases again. Therefore, in the dash-dotted brown case, the interpolated values of $\delta \theta(t)$ used when saturation is reached keep decreasing for a while, although physically this saturation should have coincided with a minimum value of $\delta \theta$. This kind of inconsistency is due to using an interpolation of the closure parameters as a function of time, which is very sensitive to any deviation in the time evolution. This is the reason why we cannot start the moment calculations too early and why we unavoidably lose convergence after some time. 

\subsubsection{An a priori closure}

\newtext{Our results show that the parameterization we introduced provides an efficient and complete framework to implement quantum closures. If one knows the “exact” closure parameters from a multi-angle calculation, then one recovers the multi-angle results with moments (dashed green curve on Figs.~\ref{fig:FFI_ee} and \ref{fig:FFI_offdiag}). This is not, however, a realistic path for moment-based neutrino flavor transformation calculations: one wants to adopt a closure \emph{a priori}, not after completing an expensive multi-angle calculation. One should expect that a closure adopted a priori will lead to some difference from a multi-angle evolution, but hopefully the difference would be `small' — this is the very same problem as in classical moment transport.

The formalism of Sec.~\ref{sec:quantum_closure} paves the way for future studies dedicated to the design of sophisticated closures, that is, accurate prescriptions for $v_P/v_E$, $\delta \theta$ and $\delta \phi$. 
For now, we will create a basic closure to show how this ideal approach could be achieved in practice. Our simple, a priori, closure is based upon the initial values of the moments and the linear phase of the instability (see Fig.~\ref{fig:FFIclosure}), which can be evaluated analytically. 

First, from the initial values for the moments given in Table~\ref{tab:initial_FFI}, the $\chi, \, \bar{\chi}$ parameters can be determined exactly and are,
\begin{equation}
    \chi = \frac{P_{zz}^{\, \widetilde{t}}}{E^{\, \widetilde{t}}} = \frac{P_{zz}^{(ee)}+P_{zz}^{(xx)}}{E^{(ee)}+E^{(xx)}} \simeq 0.339 \, ,
\end{equation}
where the pressure moments are calculated from $E$ and $F_z$ through a classical closure relation~\eqref{eq:chi_ME}. Similarly, we get $\bar{\chi} \simeq 0.352$. 

Second, during the initial linear growth phase we approximate $\theta_E \simeq \theta_P \simeq 0$ as seen in Fig.~\ref{fig:FFIclosure} (second row, left panel). Using this assumption in the definitions of $E^{\, \widetilde{z}}$ and $P_{zz}^{\, \widetilde{z}}$ [see Eq.~\eqref{eq:components_P}], we find
\begin{equation}
    \frac{v_P}{v_E} = \frac{1}{\chi} \frac{P_{zz}^{(ee)} - P_{zz}^{(xx)}}{E^{(ee)}-E^{(xx)}} \simeq 0.997 \, ,
\end{equation}
and we find similarly $\bar{v}_P/\bar{v}_E \simeq 1.005$.

The final quantities to determine are the ratio $\theta_P/\theta_E$ and the phase difference $\delta \phi$ (and likewise for antineutrinos). For these quantities we use the multi-angle linear stability analysis (LSA) method of~\cite{Froustey:2023skf}, Appendix B, with $\vec{k} = \vec{0}$ (i.e.\ the homogeneous mode) and adopt 120 angular bins. Introducing the multi-angle LSA eigenvectors $Q_\mu$ and ${\bar{Q}}_\mu$ which are defined such that $\varrho^{(ex)}(\mu) \propto Q_\mu e^{-\imath \Omega t}$ with $\Omega$ a complex frequency (and similarly for the antineutrinos), we find from the LSA that the `growth rate' $\mathrm{Im}(\Omega)$ is $\mathrm{Im}(\Omega) \simeq 2.74 \times 10^{9} \, \mathrm{s}^{-1}$ (though this is not used in defining the closure). 
The multi-angle eigenvectors are shown in  Fig.~\ref{fig:multi_eigenvec}. Note that the distributions are relatively “smooth” which suggests that a few angular moments could be sufficient to describe their main features.
\begin{figure}[ht]
    \centering
    \includegraphics[width=\columnwidth]{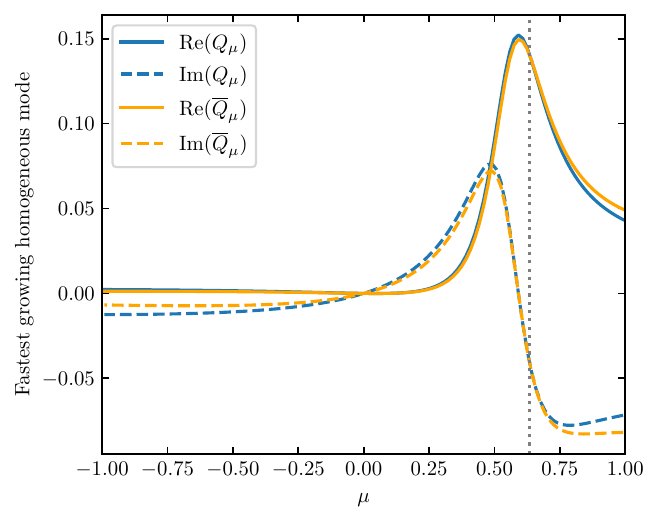}
    \caption{\newtext{Angular distribution of the multi-angle eigenvectors obtained from LSA, corresponding to the angular distribution of $\varrho^{(ex)}$ and $\bar{\varrho}^{(ex)}$ in the linear phase of the instability.}}
    \label{fig:multi_eigenvec}
\end{figure}

In the linear phase we can write the ratio of the flavor-space off-diagonal elements of the pressure and energy density moments as:
\begin{align}
    \frac{P_{zz}^{(ex)}}{E^{(ex)}} &\simeq \chi \frac{v_P}{v_E} \times \frac{\theta_P}{\theta_E} e^{-\imath (\phi_P - \phi_E)} \\
    &= \frac{\int_{-1}^{1}{\mathrm{d}\mu \, \mu^2 Q_\mu}}{\int_{-1}^{1}{\mathrm{d}\mu \, Q_\mu}} \simeq 0.552\, e^{- 0.334 \imath} \, , \nonumber
\end{align}
and using the previous results for $\chi$ and $v_P/ v_E$, we find $\theta_P/\theta_E = 1.635$ and $\delta \phi = \phi_E - \phi_P = -0.334 \, \mathrm{rad}$. A similar analysis for the antineutrinos yields $\bar{\theta}_P/\bar{\theta}_E = 1.591$ and $\delta \bar{\phi} = -0.324 \, \mathrm{rad}$. 

We now have all the ingredients for an a priori closure prescription. 
First, we consider constant values of $\{\chi, \bar{\chi}, v_P/v_E, \bar{v}_P/\bar{v}_E,\theta_P/\theta_E, \bar{\theta}_P/\bar{\theta}_E\}$ set to the values obtained from the initial conditions and LSA. We assign the \emph{sign} of the phase difference $\delta \phi$ depending on the \emph{sign} of the flavor conversion, namely:
\begin{equation}
    \begin{aligned}
        \delta \phi &= \mathrm{sgn}[\dot{E}^{(ee)}] \times 0.334 \, \mathrm{rad} \, , \\
        \delta \bar{\phi} &= \mathrm{sgn}[\dot{\bar{E}}^{(ee)}] \times 0.324 \, \mathrm{rad} \, ,
    \end{aligned}
\end{equation}
where the dot denotes the time derivative such that the closure continues to be approximately correct well after saturation (third row, right column in Fig.~\ref{fig:FFIclosure}). 

The results of the moment calculation using this prescription (“a priori closure”), along with the multi-angle results and the ones using the full interpolated closure, are shown on Fig.~\ref{fig:testclosure_FFI}.
\begin{figure}[ht]
    \centering
    \includegraphics[width=\columnwidth]{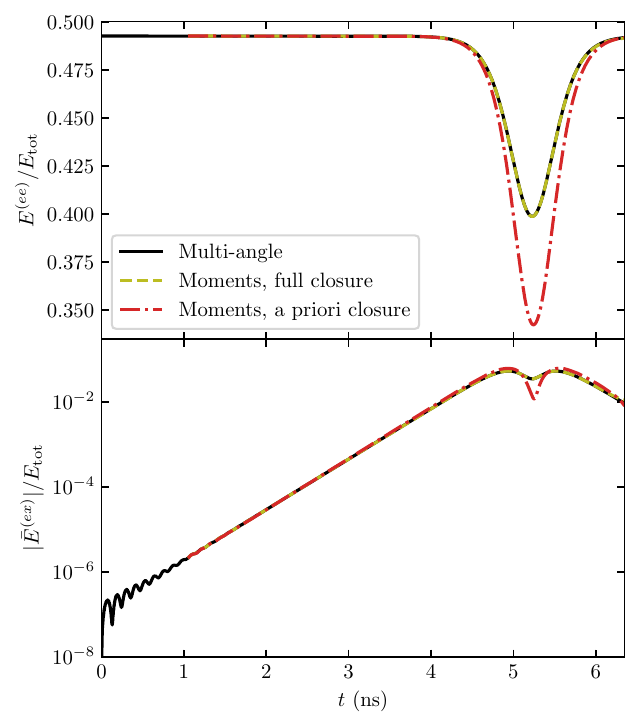}
    \caption{\newtext{Simulation of the homogeneous FFI with moments, compared with the multi-angle result. See text for details on the closure prescription. Plotting conventions are the same as Figs.~\ref{fig:FFI_ee} and \ref{fig:FFI_offdiag}.}}
    \label{fig:testclosure_FFI}
\end{figure}
Although simple, our a priori closure is seen to be extremely accurate in the linear phase of the instability, as expected (since, according to Fig.~\ref{fig:FFIclosure}, the closure parameters are then constant and consistent with the values we set). Since we do not include the variation of the closure parameters around the saturation of the instability at $t \sim 5.3 \, \mathrm{ns}$, it is not surprising that the saturation is more poorly described, with too much flavor conversion (see top panel of Fig.~\ref{fig:testclosure_FFI}). However, we emphasize that this is only a simple example of a closure prescription using our parameterization. Future work will be devoted to design better closures, which would account for the variation of all parameters and be generalized to non-homogeneous situations.}

\section{Summary and Conclusions}
\label{sec:conclusions}

In this paper we study analytic closures for two-moment neutrino quantum kinetics and derive a general parameterization of the quantum closure that can be used for quantum moment neutrino transport, a crucial path forward to tackle the computational complexity of flavor transformation in large-scale simulations. Focusing on the case of a closure relation between the energy density $E$ and pressure $P_{rr}$ moments, we provide two alternative sets of parameters that completely describe any closure [Eqs.~\eqref{eq:closure_L} and \eqref{eq:components_P}]. These parameters encode the ‘three-step’ sequence that relates the energy and pressure moments: a diagonalization of $E$, a rescaling of its eigenvalues, and finally a ‘de-diagonalization’ to obtain $P_{rr}$. We additionally prove that the closure parameters relating $E$ and $P_{rr}$ have to satisfy several constraints, related to the positive-semidefinite nature of the angular moments and the geometrical relation between them.

One of the attractive features of the two-moment formulation of kinetics is a promise that given a 'complete' closure, the evolution of the truncated set of moments is exact. We show that our parameterization satisfies this requirement on two examples, a steady-state MSW problem in the Bulb model geometry, and an homogeneous fast flavor instability in an NSM environment. By matching more and more of the closure parameters to the interpolated values from the multi-angle calculation, we are able to recover, with a moment code, the actual dynamical evolution.

Of course, this proof of concept is not generalizable as it requires the time-intensive multi-angle results to be used as input of the quantum closure. \newtext{To illustrate a path forward, we designed, in the homogeneous FFI test case, a simple closure which does not require the expensive multi-angle calculation. This closure captures some key features of the actual evolution, in spite of its simplicity.} More generally, equipped with the framework we developed, we can undertake a systematic study of closure prescriptions in the quantum case and determine, as in the classical case, the domains of applicability and limitations of various choices of quantum closures.

\begin{acknowledgments}

We are grateful to Eve Armstrong and Hiroki Nagakura for many useful discussions. 
This work was supported at NC State by DOE grant DE-FG02-02ER41216 and DE-SC00268442 (ENAF).  J.F. is supported by the Network for Neutrinos, Nuclear Astrophysics and Symmetries (N3AS), through the National Science Foundation Physics Frontier Center award No. PHY-2020275. F.F. gratefully acknowledges support from the Department of Energy, Office of Science, Office of Nuclear Physics, under contract number DE-AC02-05CH11231, from NASA through grant 80NSSC22K0719, and from the NSF through grant AST-2107932. S.R. was supported by a National Science Foundation Astronomy and Astrophysics Postdoctoral Fellowship under Award No. AST-2001760.
\end{acknowledgments}

\appendix


\section{Additional Mathematical Derivations}
\label{sec:app1}

In this appendix we collect several proofs of identities used in our paper, related to the manipulation of Hermitian matrices. Note that our foundation is slightly different than used elsewhere in the literature — e.g., \cite{Duan:2010bg}. 

\subsection{Matrix Components}
First, we define the `matrix scalar component' $M^{A}$ of a matrix $M$ with respect to the arbitrary matrix $A$ as
\begin{equation}
M^{A} = \frac{\langle A, M \rangle_{F} }{||A||^2_F}\, ,
\end{equation} 
where $\langle A, M \rangle_{F} = \mathrm{Tr}(A^\dagger M)$ is the Frobenius inner product of the two matrices, and $||A||_{F}$ is the Frobenius norm of $A$. Next, we define a basis for matrices of dimensions $R\times S$ as being a set of matrices of dimensions $R\times S$ such that any arbitrary matrix $M$ can be written as a unique linear combination of the elements of the basis with scalar coefficients. For simplicity we will use a basis such that the Frobenius norm of every element of the basis is unity and the Frobenius inner product of any element of the basis with any other element is zero. These properties match the concept of `normalization' and `orthogonality' of a set of unit basis vectors but in what follows we avoid the terms `normalized' and `orthogonal' because these terms (or similar) have different meanings when applied to matrices.  We choose the basis matrices to be Hermitian: while it is possible to use a set of non-Hermitian basis matrices\footnote{For example, one may use the equivalent of a standard vector basis where every element of the matrix is zero except for one element which is equal to 1.}, the advantage of a Hermitian basis is that the basis matrices do not change under Hermitian conjugation. In addition, if we use a set of Hermitian matrices for the basis then for an Hermitian matrix $M$, the scalar components are pure real.

For the specific case of $2 \times 2$ matrices, one possible basis is the set $\{e^{\,\widetilde{t}}, e^{\,\widetilde{x}},e^{\,\widetilde{y}},e^{\,\widetilde{z}} \}$ where $e^{\,\widetilde{t}} = \mathbbm{1}_2 /\sqrt{2}$ and $e^{\,\widetilde{x}}$, $e^{\,\widetilde{y}}$ and $e^{\,\widetilde{z}}$ the normalized Pauli matrices (i.e., the regular Pauli matrices divided by $\sqrt{2}$). We have placed a tilde over the superscripts to indicate that  $\widetilde{t}$, $\widetilde{x}$, $\widetilde{y}$, $\widetilde{z}$ are not physical space indices. 
Using the matrix basis elements $\{e^{\,\widetilde{t}}, e^{\,\widetilde{x}},e^{\,\widetilde{y}},e^{\,\widetilde{z}} \}$, the concept of matrix components allows us to write the matrix $M$ as
\begin{equation}
M = M^{\,\widetilde{t}}\,e^{\,\widetilde{t}} +  M^{\,\widetilde{x}}\,e^{\,\widetilde{x}} + M^{\,\widetilde{y}}\,e^{\,\widetilde{y}}+ M^{\,\widetilde{z}}\,e^{\,\widetilde{z}} \, ,
\end{equation}
with $M^{\,\widetilde{n}}$,  $\widetilde{n} \in\{\widetilde{t},\widetilde{x},\widetilde{y},\widetilde{z} \}$ are the scalar components. By using basis matrices with Frobenius norm of unity rather than the Pauli matrices, we avoid the additional factor of $1/2$ that one often sees when expanding a Hermitian matrix in terms of the Pauli matrices, which makes it more obvious how this expression can be easily generalized. 
In terms of the matrix elements of $M$, the scalar components of $M$ for the case where $M$ is a $2 \times2$ matrix are
\begin{equation}
\label{eq:decompo_basis}
    \begin{aligned}
    M^{\,\widetilde{t}} &= \frac{\left(M^{(11)}+M^{(22)}\right)}{\sqrt{2}} \, , \\
    M^{\,\widetilde{x}} &= \frac{\left(M^{(12)}+M^{(21)}\right)}{\sqrt{2}} \, , \\
    M^{\,\widetilde{y}} &= \imath\, \frac{\left(M^{(12)}-M^{(21)}\right)}{\sqrt{2}} \, , \\
    M^{\,\widetilde{z}} &= \frac{\left(M^{(11)}-M^{(22)}\right)}{\sqrt{2}} \, .
    \end{aligned}
\end{equation}
Note that $\mathrm{Tr}(A) = \sqrt{2}\,M^{\,\widetilde{t}}$. The four scalar components can be arranged into a 4-vector $\mathbf{M}$ 
\begin{equation}
    \begin{aligned}
    \mathbf{M} &= (M^{\,\widetilde{t}},M^{\,\widetilde{x}}, M^{\,\widetilde{y}}, M^{\,\widetilde{z}}) = (M^{\,\widetilde{t}},\vec{M}) \, .
    \end{aligned}
\end{equation}
$\vec{M}$ is the 3-vector formed by the `space' components of $M$, i.e., $\vec{M} = (M^{\,\widetilde{x}}, M^{\,\widetilde{y}},M^{\,\widetilde{z}})$. The square magnitude of $\mathbf{M}$ is $|\mathbf{M}|^2 = (M^{\,\widetilde{t}})^2 + |\vec{M}|^2 =  ||M||^2_F$ i.e. the Frobenius norm of $M$. We suggestively define the ratio of the `spatial' to `time' parts of the 4-vector as a `velocity' $\vec{v}$:  
\begin{equation}
\vec{v} = \frac{\vec{M}}{M^{\,\widetilde{t}}} \, ,
\end{equation}
and we define the `speed' $v$ as $v = |\vec{v}|$. These concepts can be extended to larger matrices by enlarging the `space' components of $M$. 

For $2\times 2$ matrices the eigenvalue matrix $\Lambda$ of $M$ is\footnote{Note that we have defined the eigenvalue matrix with the positive sign for the term proportional to $e^{\,\widetilde{z}}$. One could equally use the negative sign in the definition.}
\begin{align}
\Lambda &=  M^{\,\widetilde{t}}\,e^{\,\widetilde{t}} + |\vec{M}|\,e^{\,\widetilde{z}} \, , \label{eq:def_Lambda} \\ &= M^{\,\widetilde{t}}\,\left( e^{\,\widetilde{t}} + v\,e^{\,\widetilde{z}} \right) \, .
\end{align}
Note that for the case where $M$ is positive semi-definite, the speed is such that $0 \leq v \leq 1$. 

\subsection{Rotation and Diagonalization}
 
\paragraph*{Generic rotation ---} We now consider a general traceless, $2 \times 2$ unitary Hermitian matrix $R$ which has a 4-vector representation $\mathbf{R} = (0,\vec{R})$. The unitary property of $R$ means $R^{\dagger}\,R = \mathbbm{1}_2$ which means $|\vec{R}|^2 = 2$, and since $R$ is also Hermitian, $R^{\dagger} = R$. Therefore, $R$ is a square root of the unit matrix. If $R$ possesses these properties, we next consider how the matrix $N$ defined as $N = R M R$ is related to $M$. After writing both $M$ and $R$ in component form and using the properties of Pauli matrices, we eventually find 
\begin{equation}
\left\{
\begin{aligned}
N^{\,\widetilde{t}} &= M^{\,\widetilde{t}} \, , \\
\vec{N} &= - \vec{M} + \left( \vec{R}\cdot\vec{M} \right)\,\vec{R} \, .
\end{aligned}
\right.
\end{equation}
The second equation corresponds to the Rodrigues formula for the rotation of $\vec{M}$ by 180 degrees around the vector $\vec{R}$. 
Thus we find 
\begin{equation}
N = R\,M\,R = 2\,M^{\,\widetilde{t}}\,e^{\,\widetilde{t}} - M + \left(\vec{R}\cdot\vec{M}\right)\,R \, .
\end{equation}

\paragraph*{Diagonalization of $M$ ---} One matrix which has the required properties of $R$ is the matrix $\Upsilon$ defined to be
\begin{equation}
\Upsilon \equiv \frac{ \sqrt{2}\,\left( M + \Lambda - 2\,M^{\,\widetilde{t}}\,e^{\,\widetilde{t}}\right)}{ ||M + \Lambda - 2\,M^{\,\widetilde{t}}\,e^{\,\widetilde{t}}||_F } \, ,
\label{eq:Upsilondef}
\end{equation}
where $\Lambda$ is the eigenvalue matrix defined in Eq.~\eqref{eq:def_Lambda}.
If $M$ is Hermitian then $\Upsilon$ is also Hermitian i.e. $\Upsilon = \Upsilon^{\dagger}$. The denominator of $\Upsilon$ can be expressed in terms of the scalar components of $M$ and is
\begin{equation}
||M + \Lambda - 2\,M^{\,\widetilde{t}}\,e^{\,\widetilde{t}}||_F= \sqrt{ 2\,v\,M^{\,\widetilde{t}}\,\left(v\,M^{\,\widetilde{t}} + M^{\,\widetilde{z}} \right) } \, . 
\end{equation}
Thus we find the scalar components of $\Upsilon$ are
\begin{align} 
\Upsilon^{\,\widetilde{x}} &= \frac{M^{\,\widetilde{x}} }{ \sqrt{v\,M^{\,\widetilde{t}}\,\left(v\,M^{\,\widetilde{t}}+M^{\,\widetilde{z}}\right) }} \, , \\
\Upsilon^{\,\widetilde{y}} &= \frac{M^{\,\widetilde{y}} }{ \sqrt{v\,M^{\,\widetilde{t}}\,\left(v\,M^{\,\widetilde{t}}+M^{\,\widetilde{z}}\right) }} \, , \\
\Upsilon^{\,\widetilde{z}} &= \sqrt{ \frac{v\,M^{\,\widetilde{t}}+M^{\,\widetilde{z}} }{v\,M^{\,\widetilde{t}}}}  \, .
\end{align}
The significance of $\Upsilon$ becomes clearer if we introduce the angles $\eta$, $\theta$, $\phi$ via the definitions that 
\begin{equation}
\label{eq:RMR}
\begin{aligned}
M^{\,\widetilde{t}} &= ||M||_F\,\cos\eta \, , \\
M^{\,\widetilde{x}} &=  ||M||_F\,\sin\eta\,\sin\theta\,\cos\phi \, , \\
M^{\,\widetilde{y}} &= ||M||_F\,\sin\eta\,\sin\theta\,\sin\phi \, , \\
M^{\,\widetilde{z}} &= ||M||_F\,\sin\eta\,\cos\theta \, .
\end{aligned}
\end{equation}
In terms of these angles, we find that $\Upsilon$ has scalar components 
\begin{align}
\Upsilon^{\,\widetilde{x}} &= \sqrt{2}\,\sin\left(\frac{\theta}{2}\right)\,\cos\phi \, , \\
\Upsilon^{\,\widetilde{y}} &= \sqrt{2}\,\sin\left(\frac{\theta}{2}\right)\,\sin\phi \, , \\
\Upsilon^{\,\widetilde{z}} &= \sqrt{2}\,\cos\left(\frac{\theta}{2}\right) \, .
\end{align}
These expressions show that $\vec{\Upsilon}$ is a vector in the plane formed by $\vec{M}$ and the $\widetilde{z}$ axis which bisects $\vec{M}$ and the $+\widetilde{z}$ axis. This form for the scalar components of $\Upsilon$ also makes it easier to see that $|\vec{\Upsilon}|^2 =2$ and thus $\Upsilon$ is also unitary i.e. $\Upsilon\,\Upsilon^{\dagger} = \mathbbm{1}_2$. The purpose of the additional factor of $\sqrt{2}$ in Eq.~\eqref{eq:Upsilondef} for the definition of $\Upsilon$ was to ensure this property. Note that the determinant of $\Upsilon$ is $\mathrm{det}(\Upsilon) = -1$. 

Diagonalization of the matrix $M$ to its eigenvalue matrix $\Lambda$ can be viewed as a rotation of $\vec{M}$ around a vector until it aligns with the $+\widetilde{z}$ axis. Since the vector $\vec{\Upsilon}$ bisects $\vec{M}$ and the $+\widetilde{z}$ axis, rotating $\vec{M}$ by $180^{\circ}$ around $\vec{\Upsilon}$ will diagonalize $M$. Such a rotation is precisely described by the the transformation $M \to \Upsilon \, M \, \Upsilon$ according to Eq.~\eqref{eq:RMR}. Therefore, $\Upsilon$ \emph{is} the matrix which diagonalizes $M$, i.e.,
\begin{equation}
\Upsilon \, M\,\Upsilon = \Lambda.
\end{equation}

But there is another traceless, Hermitian and unitary matrix we can build from $M$ and $\Lambda$. This matrix we denote as $\Omega$ defined as
\begin{equation}
\Omega \equiv \frac{\sqrt{2}\left(M-\Lambda\right)}{||M-\Lambda||_F} \, .
\end{equation}
The Frobenius norm in the denominator is equal to 
\begin{equation}
||M - \Lambda ||_F= \sqrt{ 2\,v\,M^{\,\widetilde{t}}\,\left(v\,M^{\,\widetilde{t}} - M^{\,\widetilde{z}}\right)} \, ,
\end{equation}
so the scalar components of $\Omega$ are
\begin{align} 
\Omega^{\,\widetilde{x}} &= \frac{M^{\,\widetilde{x}} }{ \sqrt{v\,M^{\,\widetilde{t}}\,\left(v\,M^{\,\widetilde{t}}-M^{\,\widetilde{z}}\right) }} 
 = \sqrt{2}\,\cos\left(\frac{\theta}{2}\right)\,\cos\phi \, , \\
\Omega^{\,\widetilde{y}} &= \frac{M^{\,\widetilde{y}} }{ \sqrt{v\,M^{\,\widetilde{t}}\,\left(v\,M^{\,\widetilde{t}}-M^{\,\widetilde{z}}\right) }} = \sqrt{2}\,\cos\left(\frac{\theta}{2}\right)\,\sin\phi \, , \\
\Omega^{\,\widetilde{z}} &= \sqrt{ \frac{v\,M^{\,\widetilde{t}}-M^{\,\widetilde{z}} }{v\,M^{\,\widetilde{t}}}}= - \sqrt{2}\,\sin\left(\frac{\theta}{2}\right) \, .
\end{align}
The unitarity of $\Omega$ is easily verified and the vector $\vec{\Omega}$ is seen to be both perpendicular to $\vec{\Upsilon}$ and bisects the angle between $\vec{M}$ and the $-\widetilde{z}$ axis. So like $\Upsilon$, $\Omega$ also diagonalizes $M$ but not to $\Lambda$, instead we obtain
\begin{equation}
\Omega \, M\,\Omega = 2\,M^{\,\widetilde{t}}\,e^{\,\widetilde{t}}-\Lambda \, ,
\end{equation}
which shows that $\Omega \, M\,\Omega$ is a matrix where the $\widetilde{z}$ component of $\Lambda$ has been inverted, i.e., it now points along the $-\widetilde{z}$ axis. In other words, the two eigenvalues have been exchanged between $\Lambda$ and $2 M^{\, \widetilde{t}} e^{\, \widetilde{t}} - \Lambda$.

Finally, there is one more traceless Hermitian matrix we can build from $M$. It is given by $\Psi = \Upsilon\,\Omega$ and we find the scalar components of $\Psi$ to be 
\begin{align} 
\Psi^{\,\widetilde{x}} &= \frac{-M^{\,\widetilde{y}} }{ \sqrt{v\,M^{\,\widetilde{t}}\,\left(v\,M^{\,\widetilde{t}}-M^{\,\widetilde{z}}\right) }} = \sqrt{2}\,\sin\phi \, , \\
\Psi^{\,\widetilde{y}} &= \frac{M^{\,\widetilde{y}} }{ \sqrt{v\,M^{\,\widetilde{t}}\,\left(v\,M^{\,\widetilde{t}}-M^{\,\widetilde{z}}\right) }} = -\sqrt{2}\,\cos\phi \, ,  \\
\Psi^{\,\widetilde{z}} &= 0 \, .
\end{align}
One may easily verify that $\Psi$, like $\Upsilon$ and $\Omega$, is unitary, and that $\vec{\Psi}$ is perpendicular to $\vec{\Upsilon}$ and $\vec{\Omega}$. However $\Psi$ does not diagonalize $M$, the effect of $\Psi$ upon $M$ is to invert the spatial components of $M$ but leave the temporal component unchanged, i.e.,
\begin{equation}
\Psi \, M\,\Psi = 2\,M^{\,\widetilde{t}}\,e^{\,\widetilde{t}} -M.
\end{equation}


\subsection{Matrix Alignment}

The concept of matrix components also allows us to define the `angle' $\Xi$ between two matrices $M$ and $N$ as 
\begin{equation}
\label{eq:cosXi}
\cos \Xi = \frac{\langle M, N \rangle_{F} }{||M||_F \, ||N||_F} \, .
\end{equation}
To distinguish the `velocities', and hyperspherical angles of $M$ and $N$ we add subscripts to these quantities. In terms of these parameters we find 
\begin{align}
\cos \Xi &= \frac{ M^{\,\widetilde{t}}\,N^{\,\widetilde{t}} + \vec{M}\cdot\vec{N} }{|\mathbf{M}| \, |\mathbf{N}|} \nonumber \\
&= \frac{1 + \vec{v}_M \cdot \vec{v}_N }{\sqrt{1+v_M^2} \,\sqrt{1+v_N^2} } \\
&= \cos\eta_M\,\cos\eta_N + \sin\eta_M\,\sin\eta_N \times \nonumber \\
 &\quad \left[\cos\theta_M\,\cos\theta_N + \sin\theta_M\,\sin\theta_N\,\cos\left(\phi_M - \phi_N\right) \right] \, , \nonumber
\end{align}
showing that this definition of the angle between the matrices $M$ and $N$ is equivalent to the more familiar angle between the two 4-vectors ${\mathbf M}$ and ${\mathbf N}$ we construct from $M$ and $N$. 

Based upon this idea, we consider the more general case of the angle between the two matrices $a_M\,M + b_M\,\Lambda_M + c_M\,M^{\,\widetilde{t}}\,e^{\,\widetilde{t}}$ and $a_N\,N + b_N\,\Lambda_N + c_N\,N^{\,\widetilde{t}}\,e^{\,\widetilde{t}}$, where the two sets $\{a_M,b_M,c_M\}$ and $\{a_N,b_N,c_N\}$ are arbitrary scalar constants. 
The angle $\Theta$ between them is:

\begin{widetext}
\begin{equation}
\cos \Theta = \frac{(a_M+b_M+c_M)(a_N+b_N+c_N) \,M^{\,\widetilde{t}}\,N^{\,\widetilde{t}} + ( a_M\,\vec{M}+b_M\,\vec{\Lambda}_M)\cdot(a_N\,\vec{N}+b_N\,\vec{\Lambda}_N) }{\sqrt{(a_M+b_M+c_M)^2\,(M^{\,\widetilde{t}})^2 + |a_M\,\vec{M}+b_M\,\vec{\Lambda}_M|^2}  \, \sqrt{(a_N+b_N+c_N)^2\,(N^{\,\widetilde{t}})^2 + |a_N\,\vec{N}+b_N\,\vec{\Lambda}_N|^2}  } \, ,
\end{equation}
\end{widetext}
where we defined $\vec{\Lambda}_M$ as the spatial part of $\Lambda_M$ i.e. a vector of magnitude $|\vec{M|}$ aligned along the $\widetilde{z}$ axis. 
One can imagine defining many different angles between different matrices which can be constructed from various combinations of the coefficients $\{a,b,c\}$. The case $\{a_N = a_M=1, \, b_N = b_M=0, \, c_N = c_M=0\}$ has already been provided in Eq.~\eqref{eq:cosXi} but there are other interesting combinations. 

First, the angle between the spatial parts of $M$ and $N$ is denoted as $\xi$ which corresponds to the case where $\{a_N=a_M=1,\, b_N = b_M=0,\, c_N=c_M=-1\}$. We find:
\begin{align}
\cos \xi &= \frac{ \vec{M} \cdot \vec{N} }{ |\vec{M}| \,|\vec{N}|} \nonumber \\
   &= \frac{ \vec{v}_M \cdot \vec{v}_N }{ v_M \, v_N} \\
   &= \cos\theta_M\,\cos\theta_N + \sin\theta_M\,\sin\theta_N\,\cos\left(\phi_M - \phi_N\right) \, , \nonumber \\ \nonumber \\ \nonumber
\end{align}
and we observe that $\Xi$ and $\xi$ are related via 
\begin{equation}
\begin{aligned}
\cos \Xi &= \frac{1+v_M\,v_N\,\cos\xi}{\sqrt{1+v_M^2}\,\sqrt{1+v_N^2}} \\
&= \cos\eta_M\,\cos\eta_N + \sin\eta_M\,\sin\eta_N\,\cos\xi \, .
\end{aligned}
\end{equation}
The angle $\Gamma$ between the two eigenvalue matrices $\Lambda_M$ and $\Lambda_N$ corresponds to the case $\{a_N=a_M=0, \, b_N = b_M = 1, \, c_N=c_M=0\}$, for which:
\begin{align}
\cos \Gamma &= \frac{ M^{\,\widetilde{t}}\,N^{\,\widetilde{t}} + |\vec{M}|\,|\vec{N}| }{|\mathbf{M}| \, |\mathbf{N}|} \nonumber \\
&= \frac{1 + v_M \,v_N }{\sqrt{1+v_M^2} \,\sqrt{1+v_N^2} } \\
&= \cos\eta_M\,\cos\eta_N + \sin\eta_M\,\sin\eta_N \, . \nonumber
\end{align}
From these results we see that the angles $\Xi$, $\xi$ and $\Gamma$ are related via 
\begin{equation}
\cos\xi = \frac{\cos\Xi - \cos\eta_M\,\cos\eta_N}{\cos\Gamma - \cos\eta_M\,\cos\eta_N} \, .
\label{eq:cosxiidentity}
\end{equation}

\bibliography{main}

\end{document}